\documentclass[a4paper,11pt]{article}
\usepackage{jheppub}
\usepackage{dsfont}

\renewcommand{\i}{\mathrm{i}}
\renewcommand{\d}{\partial}

\usepackage[normalem]{ulem}
\usepackage[dvipsnames]{xcolor}
\usepackage{comment}

\begin{document}

\title{Holographic Soliton Crystals for Dense Nuclear Matter and Neutron Stars}

\author[a]{Lorenzo Bartolini,}
\author[b,c]{Sven Bjarke Gudnason,}
\author[a]{Jonas Mager,}
\author[a]{Anton Rebhan}
\affiliation[a]{Institut f{\"u}r Theoretische Physik, Technische Universit{\"a}t Wien,\\
Wiedner Hauptstrasse 8-10, A-1040 Vienna, Austria}
\affiliation[b]{Institute of Contemporary Mathematics, School of
  Mathematics and Statistics, Henan University, Kaifeng, Henan 475004,
  P.~R.~China}
\affiliation[c]{Department of Physics, Chemistry and Pharmacy,
  University of Southern Denmark, Campusvej 55, 5230 Odense M,
  Denmark}

\abstract{We construct a nuclear-matter equation of state (EOS) from holographic QCD in the Witten--Sakai--Sugimoto (WSS) model, going beyond the homogeneous ansatz by building dense baryonic matter more directly from the solitonic description of holographic baryons. Employing the two-baryon interaction potential obtained from the linearized soliton tails sourced in the curved background by point-like (core) instantons, we assemble an infinite face-centered cubic (FCC) crystal with nearest neighbors in the most attractive channel as an approximation to the quantum liquid of baryons and compute the energy density $\mathcal{E}(n_B)$, the chemical potential $\mu_B$, and the pressure $P$.
We calibrate the symmetric-matter EOS by fixing the 't Hooft coupling $\lambda$ and the WSS scale $M_{\rm KK}$ to saturation-density and onset-chemical-potential properties, and by including a quark-mass term to reproduce the physical pion mass $m_\pi=135\,\mathrm{MeV}$. 
This fit turns out to remain reasonably close to the parameters required
by the vacuum meson sector, while their modification is consistent with Brown-Rho scaling in a dense medium, and the incompressibility at saturation is of the correct order of magnitude, both in clear contrast to the homogeneous approximation. We then extend to beta-equilibrated matter, using phenomenological input for the symmetry energy, and obtain hybrid EOS and neutron-star observables that are compatible with the NICER constraints.}


\maketitle

\flushbottom

\section{Introduction and state-of-the-art}
\label{sec:intro}

Neutron stars occupy a fascinating intersection of gravitational physics, nuclear physics, astrophysics, and neutrino (particle) physics.
The observation of neutron star (NS) mergers through gravitational waves and new radius-mass measurements of stable neutron stars places
stringent bounds on nuclear physics' equations of state 
as well as tidal deformability, for reviews see Refs.~\cite{Burgio:2021vgk,Jarvinen:2021jbd,Hoyos:2021uff}. 
The largest part of the mass of the neutron star is due to the neutrons whose mass is entirely a non-perturbative effect of quantum chromodynamics (QCD), which unfortunately is unable to give analytic predictions for the nucleon -- since perturbation theory breaks down.
Lattice QCD can provide good results, but only at small or vanishing chemical potential due to a technical issue known as the sign problem \cite{deForcrand:2009zkb}.

Holography refers to a broad class of theories possessing a duality between a gravitational theory in $D+1$ dimensions and a gauge theory in $D$ dimensions, with Maldacena's discovery of $\mathcal{N}=4$ super-Yang-Mills (sYM) providing the AdS$_5$/CFT$_4$ duality as a prime example \cite{Maldacena:1997re}.
Although it was known long before Maldacena's 1997 paper that the zero modes 
coincide between the two sides of the duality, the ground-breaking idea leading to the full duality was the conjecture that all massive states also match on both sides of the duality: In the case of $\mathcal{N}=4$ sYM, this is demonstrated by matching the primary CFT operators on each side.
The power of the duality is due to the gravitational theory at weak coupling being mapped to a gauge theory at strong coupling -- especially enticing for QCD, if, as conjectured, such dualities hold more generally.
The broad class of holographic models attempting to describe low-energy QCD at strong coupling is known as ``holographic QCD'' and comprises bottom-up variants such as the hard-wall \cite{Polchinski:2001tt,Boschi-Filho:2002wdj} and soft-wall models \cite{Karch:2006pv} like V-QCD \cite{Gursoy:2007cb,Jarvinen:2011qe}, but also in the theoretically more appealing top-down models based on concrete string-theoretic constructions, such as the Witten-Sakai-Sugimoto (WSS) model \cite{Witten:1998zw,Sakai:2004cn}.

The low-energy effective description of the WSS model is given in terms of a 5-dimensional gauge theory containing the Yang-Mills (YM) term, the Chern-Simons (CS) term, and, for massive pions, the Aharony-Kutasov (AK) term \cite{Aharony:2008an} with the latter being a Wilson line connecting the left- and right-handed flavor branes, corresponding to left- and right-handed quarks.
The baryon, or specifically the nucleon, in the WSS model is described by an instanton-like solution also called the Sakai-Sugimoto (SS) soliton \cite{Bolognesi:2013nja}.
Integrating over the holographic direction, spanning from the infrared (IR) to the ultraviolet (UV), reduces the 5-dimensional gauge theory to the Skyrme model coupled to an infinite tower of vector mesons \cite{Sakai:2004cn,Sutcliffe:2010et}.
This link between holographic ``bulk'' theories and the Skyrme model actually predates the holographic literature, as it was understood mathematically by Atiyah and Manton that the Skyrme field is given by the holonomy of the instanton gauge field \cite{Atiyah:1989dq}.
The ``bulk'' geometry then fixes every coefficient of interaction between the pions and all the vector mesons in terms of integrals over the holographic direction \cite{Sakai:2004cn,Sutcliffe:2010et}.
This mechanism \cite{Ma:2012kb} is directly related to hidden local symmetry \cite{Bando:1987br,Harada:2003jx}.

The Skyrme model predicts properties of nuclei \cite{Adkins:1983ya}, but gives rise to too large binding energies \cite{Manton:2022fcb}. Phenomenologically reasonable binding energies can be obtained by including two or three, or preferably a whole tower of, vector mesons \cite{Sutcliffe:2011ig}.
Finite density predictions with the chiral solitons or the Skyrmion lattices \cite{Klebanov:1985qi,Goldhaber:1987pb} lead to face-centered-cubic (FCC) lattice structure near and above saturation densities \cite{Kugler:1988mu,Kugler:1989uc,Adam:2023cee} and is predicted to form body-centered-cubic (BCC) lattices for $n\gtrsim3n_S$.
The FCC lattice is a lattice of Skyrmions, whereas the BCC lattice is a lattice of half-Skyrmions \cite{Manton:1994rf}, each corresponding to a twisted instanton in a cell with (anti-)periodic boundary conditions due to 't Hooft \cite{tHooft:1979rtg,tHooft:1981nnx}.
In the WSS model, it has been conjectured \cite{Rho:2009ym} that together with a restoration of chiral symmetry the BCC phase also takes place at $n\gtrsim3n_S$, dubbed a phase of dyonic salt, as the half-instantons are also called dyons.
This dyon is half an instanton (hence with $\pi_3(S^3)\subset[T^4,B_{SO(3)}]\ni N_B=1/2$)\footnote{Fractionally charged solitons are well known to physicists, but they are also mathematically precise: Indeed, (minus) the second Chern number for an $SU(N_f)/\mathbb{Z}_{N_f}$ gauge theory is only an integer without twists and on $T^4$ is given by $[T^4,B_{SO(3)}]\ni N_B=\frac{1}{64\pi^2}\int d^3x\,dz\,\epsilon^{MNPQ}F_{MN}^aF_{PQ}^a=\nu+\frac{N_f-1}{8}\epsilon^{MNPQ}n_{MN}n_{PQ}$, where $\nu\in\mathbb{Z}$ is the usual integer part, $n_{MN}\in\mathbb{Z}$ are integer twists and $N_f=2$ in this paper \cite{vanBaal:1982ag,Nash:1982kp}.
We can intuitively understand this result as the gauge fields transform trivially under the center of $SU(N_f)$ and therefore the homotopy should be computed for the gauge group $SU(N_f)/\mathbb{Z}_{N_f}$, which for $N_f=2$ yields the gauge group $SO(3)$.
The complete classification of the homotopy is done only with $K$-theory \cite{Nash:1982kp}, but a simplistic argument explains that the radius of $SU(2)$ is two times that of $SO(3)$ and hence the topological charge is $\frac12\mathbb{Z}$ on a four-torus.
On $\mathbb{R}^4$ this does not work, since the point-compactification of space to $S^3$ does not allow for twists -- the point at infinity is a single point.
} and has monopole ($\pi_2$) charge $\pm1$. 
A full unit-charge instanton can thus be viewed as a bi-dyon or a dyon pair with total monopole charge zero, but naturally a full unit instanton charge.

Progress regarding finite-density computations using the Skyrme-model, coupled with the $\rho$ and the $\omega$ vector mesons \cite{Ma:2013ooa}, has been made in the literature leading to good NS phenomenology \cite{Ma:2019ery}, which surprisingly leads to a fast convergence of the speed-of-sound to the conformal value above $n\gtrsim3n_S$ implying the emergence of a conformal symmetry and illustrating the importance of the possible transition to half-Skyrmions.
In the Skyrme model with the sextic derivative term\footnote{The sextic derivative term is included in the WSS model in terms of the vector mesons, which can be seen by integrating out the lowest vector meson properly \cite{Bartolini:2017sxi}. The coupling of the sextic term, after integrating out the vector mesons, is, however, fixed by the geometry and is proportional to $N_c/\lambda$.
}, a BPS-limit of the model has been utilized to describe the dense matter EOS for neutron stars, giving reasonable results \cite{Adam:2014dqa} but depending drastically on the pion potential used.
Although this BPS-limit of the model gives rise to a version of the liquid droplet model of nuclei and allows for solving the Einstein equation without the mean-field approximated Tolman-Oppenheimer-Volkoff (TOV) equations \cite{Adam:2015lpa}, it lacks low-density physics connecting smoothly to the meson sector of the model, which was then added by means of a phenomenological crust \cite{Adam:2020djl} enlarging the already too-large neutron stars.
Interpolating between the BPS-model and the Skyrme crystal leads to good agreement with the phenomenological constraints on neutron stars \cite{Adam:2020yfv}, thus implying the importance of the crystalline structure in neutron star phenomenology \cite{Adam:2021gbm,Adam:2022aes}, for a review see Ref.~\cite{Adam:2023cee}.
The extension of the Skyrme model to three flavors, thus including strange degrees of freedom, has been carried out, giving rise to a first-order phase transition to a kaon-condensed phase \cite{Adam:2022cbs}.
Finally, in the Skyrme model the isospin density -- corresponding to the neutron/proton relative density -- has currently been taken into account perturbatively by means of zero-mode quantization following the seminal work of Adkins-Nappi-Witten (ANW) \cite{Adkins:1983ya}.

In the WSS model, the equivalent zero-mode quantization can be taken into account (again following ANW) but it is indeed equivalent to a chemical potential that is usually employed in the holographic dictionary \cite{Bartolini:2023eam} -- albeit a technical issue is that a boundary term must be dropped from the CS term to make the equivalence between the rigid-body isospin zero-mode quantization and the chemical potential imposed at the holographic boundary.

Although it has been possible to construct lattices and explicit numerical computations in a unit cell with periodic boundary conditions in the case of the Skyrme model with or without a finite number of vector mesons, the numerical state-of-the-art for the WSS model is limited by the fact that not just a 2-by-2 scalar field in 3 dimensions must be computed: instead a 2-by-2 five-vector gauge field yielding 20 nonlinear equations in 4 spatial and curved dimensions.
These equations can, in the static case, be reduced to 13 nonlinear equations in 4 spatial, curved dimensions by assuming that $A_0=0$ and $\widehat{A}_{i,z}=0$.
This numerical challenge has led to the introduction of the homogeneous ansatz in the WSS model. 
Although the idea was first eliminated by a no-go theorem \cite{Rozali:2007rx}, a trick of introducing a discontinuous field configuration led to an explicit and simple construction with instanton or baryon number proportional to the size of the discontinuity of the field at the IR end of the geometry -- also called the ``tip of the cigar''.
The advantage of the homogeneous ansatz is clear: 13 coupled nonlinear partial differential equations (PDEs) become 2 coupled nonlinear ordinary differential equations (ODEs). The degrees of freedom were reduced from 52 to 2.

Indeed, the homogeneous ansatz has been an important stepping stone, enabling explicit computations both in the WSS model and in bottom-up models to be carried out, and with some partial success. 
Application of the homogeneous ansatz to the WSS model allowed for the calculation of a first-order onset of baryonic matter \cite{Li:2015uea}, symmetry energy and isospin dependence of nuclear matter \cite{Kovensky:2021ddl,Bartolini:2022gdf}, neutron star matter with resolved composition \cite{Kovensky:2021kzl,Kovensky:2021wzu,Bartolini:2023wis} and the phase diagram in presence of meson condensation \cite{Kovensky:2023mye}. Within the bottom-up hard-wall model the homogeneous ansatz was used to derive a first-order baryonic onset and the phase diagram \cite{Bartolini:2022rkl}, neutron star matter \cite{Bartolini:2022rkl,Fujii:2025umi}, hybrid EOS for neutron star matter \cite{Wang:2025tyd}, and symmetry energy \cite{Bartolini:2022gdf}. It was also used in the VQCD model to compute hybrid EOS for neutron star matter \cite{Jokela:2020piw}, symmetry energy and isospin dependence \cite{Bartolini:2025sag}, gravitational wave spectrum from neutron stars merger events \cite{Ecker:2019xrw}, and locating the QCD critical end point from neutron star data \cite{Ecker:2025vnb} (see also Ref.~\cite{Jarvinen:2021jbd} for a review).
\footnote{
For other holographic QCD studies of dense baryonic matter for neutron stars, see also
e.g.~Refs.~\cite{Ghoroku:2013gja,Pinkanjanarod:2020mgi}
for the D4/D8/$\overline{\textrm{D8}}$ model, \cite{Hoyos:2016zke,BitaghsirFadafan:2020otb} for the D3/D7 model, \cite{Zhang:2022uin,Liu:2024efy} the for Einstein-Maxwell-Dilaton model, and \cite{Ghoroku:2021fos,Ghoroku:2023uxr} for a color-superconducting Yang-Mills model with an instanton gas.
}
Nevertheless, the model with two free parameters, i.e.~the mass scale $M_{\rm KK}$ and the 't Hooft coupling ($\lambda$), is unable to give rise to the continuous matching of energy density, pressure, and chemical potential at a given density (say an order-one number times the saturation density) with phenomenological values.
This is mathematically speaking not surprising: one cannot fit 3 quantities with 2 parameters. One could hope for some kind of miracle that the WSS model would give rise to the physically correct third quantity when the other two were fitted: but this is not the case, as the chemical potential at saturation density is generally speaking much too large compared with the nucleon mass, when the energy density and the pressure are fitted to phenomenological values (i.e.~by adjusting $M_{\rm KK}$ and $\lambda$).
A recent attempt at improving the homogeneous ansatz in the WSS model has been made in Ref.~\cite{Ecker:2025sjb}, where more than one discontinuity in the holographic direction has been introduced. However, the compression modulus still turns out to be an order of magnitude too large with respect to the phenomenological upper bound of about 300 MeV. 

It is well-known that there is a laundry list of problems with the WSS model as it is based on an inexact duality in the large-$N_c$ (number of colors) and large-'t Hooft coupling limits.
However, in this work we suggest that the homogeneous ansatz is not capturing the phenomenology of baryons at the densities relevant for neutron stars and propose that a better approximation is provided by an alternative crystal-like construction, summing the energy contributions from the holographic two-instanton interaction potential over all relevant neighbors in an FCC lattice.
Although the interaction potential is based on the soliton tail interaction with another instanton as a source -- and hence is intrinsically limited to asymptotically large distances -- it has been shown, in the case of the Skyrme model, that this interaction potential extrapolates remarkably well to shorter distances than is mathematically expected \cite{Gudnason:2020arj}. 
Using this method and including also the pion mass parameter, we can compute the isospin symmetric equation of state (EOS) for nuclear matter at not too large densities (ignoring the possibility of a transition to the dyonic BCC phase discussed in Ref.~\cite{Rho:2009ym}),
which covers most of the range relevant for neutron-star physics. While at densities higher than $3n_S$ the holographic realization of the transition to half-Skyrmions is expected to take place, our numerical results imply that the densities at the center of the average neutron stars never reach these values. On the other hand, at high densities, the many approximations in our treatment of the WSS model break down anyway: while this
precludes the precise determination of its predictions for the heaviest stable neutron stars (such as their mass $M_{\rm TOV}$), our present work focuses on connecting holographic nuclear matter with phenomenology around saturation, which is insensitive to the presence of other phases at higher density.

Phenomenological EOSs are, however, not isospin symmetric: the number of neutrons is generally larger than the number of protons -- hence the name \emph{neutron} star!
In this work, and due to the construction of the energy density from an FCC crystal of instantons, we take the isospin contribution to the energy density as a phenomenological input: we incorporate the beta-equilibrated nuclear matter energy contribution fitted to the physically measured symmetry energy at saturation density.
Below saturation density, electromagnetic interactions and exotic phases such as nuclear pasta become important: these inhomogeneous phases can in principle be treated within holographic models, and the first realization of a holographic NS crust was attempted within the WSS model in Ref.~\cite{Kovensky:2021kzl}, where it is described as a mixed phase of bubbles of homogeneous nuclear matter and leptons. The construction of the mixed phase requires the calculation of the symmetry energy and the surface tension of the domain wall that separates the nuclear matter bubbles from the gas of leptons, both of which are not yet available. For this reason, below saturation density, we simply switch to the EOS provided by nuclear EFT.

Above saturation density, our description assumes a single phase of
nuclear matter described by a densest sphere packing of SS solitons instead of
a homogeneous field configuration. Including the kinetic energy inherent in
a quantum mechanical description, but suppressed in the present treatment
as a $1/N_c$ correction, should turn the crystal into a fluid. The particular crystalline structure of our ansatz (FCC) and
its symmetries 
should thus not be expected to be realized beyond microscopic scales;
our particular inhomogeneous ansatz should rather be considered just as the attempt towards 
a more faithful description of the
interactions of dense holographic baryons in the regime where their
spatial separation is larger than the size of their cores. However, as pointed out in Ref.~\cite{Kugler:1989uc}, experience with standard liquid-solid transitions shows that, even in a liquid, short-range order reminiscent of the solid phase can be assumed to be retained.

The paper is organized as follows.
In Sec.~\ref{sec:model} we review the WSS model and set up our notations.
In Sec.~\ref{sec:homogeneous} we illustrate the issue with the homogeneous ansatz, and in Sec.~\ref{sec: nuclear matter} we review the crystal construction of instantons in the WSS model.
In Sec.~\ref{sec:symmetricfit} we introduce our fit to phenomenological data for the symmetric (part of the) nuclear matter, and in Sec.~\ref{sec:betaeq} the phenomenological input describing the symmetry energy and beta equilibrium is explained. Finally in this section we apply the construction to neutron star phenomenology and obtain good results consistent with the newest NICER constraints.
We conclude with a discussion and outlook in Sec.~\ref{sec:conclusions}.
We have relegated some details to the appendices: App.~\ref{app:equalfooting} contains a direct comparison between the homogeneous ansatz and the FCC crystal EOSs.
The FCC shell counting is given in App.~\ref{app:FCC-shell-counting}, and a comparison between the simple cubic (SC) and FCC crystal structures is given in App.~\ref{app:latticegeometry},
demonstrating that our main conclusions do not depend
too much on the particular crystalline structure assumed in
our inhomogeneous set-up.

\section{The WSS model and baryons}\label{sec:model}

The WSS model describes the flavor sector of large-$N_c$ QCD by means of probe D8/$\overline{\mathrm{D8}}$ branes embedded in the D4 background compactified on a Kaluza-Klein circle which in the low-energy limit reduces the dual theory to non-supersymmetric Yang-Mills theory \cite{Witten:1998zw,Sakai:2004cn,Sakai:2005yt}. In this limit, the open-string modes on the flavor branes reduce to a five-dimensional $U(N_f=2)$ Yang-Mills-Chern-Simons theory on a curved background. In the conventions used throughout this work, the (flavor) gauge field is decomposed as
\begin{equation}
    \mathcal{A}_\alpha = A_\alpha + \widehat{A}_\alpha\frac{\mathds{1}}{2},
    \qquad A_\alpha = A_\alpha^a\frac{\sigma^a}{2},
    \qquad \alpha=0,M, \qquad M=i,z, \qquad i=1,2,3,
\end{equation}
and the effective action reads
\begin{align}
    S &= S_{\rm YM}+S_{\rm CS},\\
    S_{\rm YM} &= -\kappa M_{\rm KK}\int d^4x\,dz\;\text{tr}\left[\frac{1}{2}h(z)\,\mathcal{F}_{\mu\nu}^2+k(z)\,\mathcal{F}_{\mu z}^2\right],\\
    S_{\rm CS} &= \frac{N_c}{24\pi^2}\int_{M^4\times\mathbb{R}}\omega_5(\mathcal{A}),
\end{align}
with
\begin{equation}
    \kappa=\frac{\lambda N_c}{216\pi^3},
    \qquad h(z)=\left(1+M_{\rm KK}^2z^2\right)^{-1/3},
    \qquad k(z)=1+M_{\rm KK}^2z^2,
\end{equation}
and $z \in (-\infty,\infty)$. The Chern-Simons term is given by
\begin{equation}
    \omega_5(\mathcal{A})=\text{tr}\left(\mathcal{A}\mathcal{F}^2-\frac{\i}{2}\mathcal{A}^3\mathcal{F}-\frac{1}{10}\mathcal{A}^5\right),
    \qquad \mathcal{F}=d\mathcal{A}+\i\mathcal{A}\wedge\mathcal{A}.
\end{equation}

The only dimensionful parameter of the model is $M_{\rm KK}$, which sets the overall energy scale, and the only other free parameter is the dimensionless 't Hooft coupling $\lambda$.
Working in units of $M_{\rm KK}$ effectively sets $M_{\rm KK}:=1$ in the equations and then lengths and the gauge fields are given in units of $M_{\rm KK}^{-1}$.

Note that we are considering the original version of the WSS model, with
antipodal positions of the D8/$\overline{\mathrm{D8}}$ branes and confining
geometry, which has proved to provide a surprisingly good description of low-energy meson physics with its minimal set of free parameters \cite{Sakai:2004cn,Sakai:2005yt,Hechenberger:2023ljn}. Other studies of the WSS model such as Ref.~\cite{BitaghsirFadafan:2018uzs} have employed a smaller D8-$\overline{\mathrm{D8}}$ brane separation $L$ such that chiral symmetry breaking can also take place in a deconfined geometry, where $M_{\rm KK}$ and thus
the deconfinement temperature can be sent to zero while keeping $L$ fixed.

Baryons are instanton-like solitons of the $SU(2)$ part of the five-dimensional gauge field \cite{Hata:2007mb,Hashimoto:2008zw}. Their baryon number is the instanton number in the four-dimensional space spanned by $(\vec{x},z)$,
\begin{equation}
\label{eq:baryonnumber}
    N_B = \frac{1}{64\pi^2}\int d^3x\,dz\;\epsilon^{MNPQ} F_{MN}^a F_{PQ}^a. 
\end{equation}

At large $\lambda$, the soliton is localized near the tip of the cigar \cite{Witten:1998zw} at $z=0$, and its size scales as $\rho\sim\lambda^{-{1}/{2}}$. In the instanton core, the geometry is thus approximately  flat and one can use the Belavin-Polyakov-Schwartz-Tyupkin (BPST) solution \cite{Belavin:1975fg}, supplemented by the abelian Coulomb field sourced by the Chern-Simons term \cite{Hata:2007mb}:
\begin{equation}
\begin{aligned}
\label{eq:BPSTplusabelian}
    A_M^{\rm cl} &= -\i f(\xi)\,g\,\partial_M g^{-1},
    \qquad \widehat{A}_0^{\rm cl} = \frac{N_c}{8\pi^2\kappa}\frac{1}{\xi^2}\left[1-\frac{\rho^4}{(\rho^2+\xi^2)^2}\right],\\
    A_0^{\rm cl} &=0,
    \qquad \widehat{A}_M^{\rm cl}=0,
\end{aligned}
\end{equation}
where
\begin{equation}
    f(\xi)=\frac{\xi^2}{\xi^2+\rho^2},
    \qquad g(x)=\frac{(z-Z)-\i(\vec{x}-\vec{X})\cdot\vec{\sigma}}{\xi},
    \qquad \xi^2=(z-Z)^2+|\vec{x}-\vec{X}|^2.
\end{equation}
The leading curvature correction induces a potential for the collective coordinates $\rho$ and $Z$,  whose minimum lies at \cite{Hata:2007mb}
\begin{equation}
  Z_{\rm cl}=0, \qquad   \rho_{\rm cl}^2 = \frac{N_c}{8\pi^2 \kappa}\sqrt{\frac{6}{5}}.
\end{equation}

The flat-space BPST solution provides a good approximation of the center of the instanton at large $\lambda$, where $\rho$ is small, but not of its asymptotic behavior far from the core. To understand the long-distance fields, one must instead solve the linearized Yang-Mills equations in the curved WSS background \cite{Hashimoto:2008zw}. To do so, we introduce Green's functions
\begin{align}
    \mathcal{G}(\vec{x},z;\vec{X},Z) &=  \sum_{n=1}^{\infty}\frac{\psi_n(z)\psi_n(Z)}{c_n}Y_n(|\vec{x}-\vec{X}|),\\
    \mathcal{H}(\vec{x},z;\vec{X},Z) &=  \sum_{n=0}^{\infty}\frac{\phi_n(z)\phi_n(Z)}{d_n}Y_n(|\vec{x}-\vec{X}|),\\
    Y_n(r)&=-\frac{1}{4\pi}\frac{e^{-k_n r}}{r},
\end{align}
where $\psi_n$ and $\phi_n$ are holographic wave functions of the mesonic modes,\footnote{Note that for $\phi_n(z),\psi_n(z)$ we follow the normalization from Ref.~\cite{Baldino:2017mqq}, which differs from that employed in Ref.~\cite{Hashimoto:2008zw}, hence the different overall factors in the definitions of $\mathcal{G},\mathcal{H}$.} which obey
\begin{equation}
\begin{aligned}
\label{eq:modeeq}
    -h(z)^{-1}\d_z\left( k(z)\d_z\psi_n(z) \right) &= k_n^2 \psi_n(z), \qquad  \int dz \; h(z) \psi_n \psi_m = c_n \delta_{nm},\\
    \phi_n(z)&= \d_z\psi_n(z),\qquad \int dz \;k(z) \phi_n \phi_m = d_n \delta_{nm},
\end{aligned}
\end{equation}
where $k_n$ denotes the corresponding particle masses.
The mode sum for the function $\mathcal{H}$ includes the derivative $\phi_0(z)$ of a non-normalizable mode $\psi_0(z)$ corresponding to the pion, given by:
\begin{equation}
    \phi_0(z)= \frac{1}{k(z)}, \qquad d_0=\pi.
\end{equation}
The static tails of a soliton centered at $(\vec{X},Z)$ can then be written as
\begin{align}
\label{eq:farfield}
    \widehat{A}_0^{\rm tail} &\simeq -\frac{N_c}{2\kappa}\,\mathcal{G}(\vec{x},z;\vec{X},Z),\\
    \label{eq:farfield2}
    A_i^{\rm tail} &\simeq -2\pi^2\rho^2\,\sigma_a \left(\epsilon_{iaj}\frac{\partial}{\partial X_j}-\delta_{ia}\frac{\partial}{\partial Z}\right)\mathcal{G}(\vec{x},z;\vec{X},Z),\\
    \label{eq:farfield3}
    A_z^{\rm tail} &\simeq -2\pi^2\rho^2\,\sigma_a\frac{\partial}{\partial X_a}\mathcal{H}(\vec{x},z;\vec{X},Z).
\end{align}
In the overlap region $\rho\ll\xi\ll1$, these linearized tails match smoothly onto the BPST core. This matched core-tail description is the main ingredient used in section \ref{sec: nuclear matter} to construct the two-baryon potential.

The soliton, described by Eq.~\eqref{eq:BPSTplusabelian} near the core and by Eqs.~\eqref{eq:farfield}, \eqref{eq:farfield2} and \eqref{eq:farfield3} at large distances, has the so-called standard orientation in $SU(2)$. Any other $SU(2)$ orientation can be reached by the transformation on the gauge field $\mathcal{A}\rightarrow B \mathcal A B^\dagger$, where $B\in SU(2)$ is constant. In the lattice constructions in Sec.~\ref{sec: nuclear matter}, the relative orientation of neighboring nucleons  plays an important role.

To account for the short-range nature of the nuclear interactions, it is important to include a non-zero pion mass in the model. This can be done by considering long open strings, which, as described in Ref.~\cite{Aharony:2008an}, are dual to the quark-mass term. This induces the additional term\footnote{We report here only the effective action after reduction to four dimensions: the full string-theoretic operator is obtained from the action of a fundamental string stretched between the $\rm D8/\overline{D8}$-branes. The Nambu-Goto part of the action contributes to the overall constant $c$, and the holonomy of $\mathcal{A}_z$ is the interaction between the string's endpoints and the brane's world-volume.} 
\begin{equation}
\label{eq:AK}
    S_{\rm AK}= c\int d^4x \;\text{tr}\,\mathcal{P}\left[\left(M e^{-\i\int dz \mathcal{A}_z} - \mathds{1} \right) + \text{h.c.}\right],\qquad c=\frac{\lambda^\frac{3}{2}}{3^{\frac{9}{2}}\pi^3},
\end{equation}
in the effective four-dimensional action.
For symmetric matter, we assume degenerate light quark masses $m_u=m_d=m$, so that the quark mass matrix is $M=m\mathds{1}$.
This term, when the exponential is expanded to second order, furnishes the field $\mathcal{A}_z$ with a mass term in its equations of motion \cite{Baldino:2021uie}:
\begin{equation}
   \kappa k(z)\left(\d_i\d_i A_z -\d_i \d_zA_i\right) -mc \int_{-\infty}^{+\infty}dz' A_z(x,z') = \text{Source terms}.
\end{equation}
The parameters $\kappa,m$ then set the values for the mass of the pion multiplet:
\begin{equation}
    k_0^2 = m_\pi^2= \pi a, \qquad a\equiv \kappa^{-1}mc = \frac{8\lambda^{\frac{1}{2}}}{3^{\frac{3}{2}}N_c}m.
\end{equation}
For each choice of the two free parameters $\lambda$ and $M_{\rm KK}$, we fit $m$ to yield the physical pion mass $m_\pi=135\,\mathrm{MeV}$.
The presence of $S_{\rm AK}$ also induces a deformation of the instanton, which is now stabilized at a smaller size:
\begin{equation}
   \rho = \rho_{\rm cl}\left(1-\frac{36 \mathcal{I}}{\sqrt{\pi}}\left(\frac{6}{5}\right)^{\frac{1}{4}}\frac{m}{N_c} +\mathcal{O}\left(\left(\frac{m}{N_c}\right)^2\right)\right),
\end{equation}
with $\mathcal{I}\approx 1.104$ defined in Eq.~\eqref{calI}.

\section{Smeared holographic nuclear matter}\label{sec:homogeneous}

The description of dense nuclear matter requires an understanding of interacting multi-soliton solutions, a formidably challenging task. The analytic and numerical difficulties have led many authors to consider a simplified setup, where baryons are uniformly smeared over $\mathbb{R}^3$, the so-called homogeneous ansatz,
in the WSS model \cite{Rozali:2007rx,Ghoroku:2012am,Li:2015uea,Preis:2016fsp,Kovensky:2021ddl,Kovensky:2021kzl,Kovensky:2021wzu,Bartolini:2022gdf,Bartolini:2023wis,Kovensky:2023mye} as well as in various bottom-up models \cite{Jokela:2018ers,Ishii:2019gta,Jokela:2020piw,Jokela:2021vwy,Bartolini:2022rkl,Jarvinen:2023jbr,Bartolini:2025sag}.
This approximation is implemented by assuming that the gauge fields depend only on the holographic coordinate $z$ and are independent of $x^\mu$. In this way, the underlying many-soliton solution is effectively replaced by a one-dimensional system, which makes the study of dense matter technically much more tractable.

In the homogeneous ansatz, the gauge field is parametrized by two functions $H(z),\hat a_0(z)$ as\footnote{Here we assume symmetric nuclear matter, thus setting the isospin chemical potential to zero.}
\begin{equation}
\label{eq:homansatz}
    \mathcal{A}_i = -\frac{H(z)}{2} \sigma^i\qquad,\qquad \mathcal{A}_0=\frac{\widehat{a}_0(z)}{2}\mathds{1}.
\end{equation}
To allow for a non-vanishing baryon number \eqref{eq:baryonnumber},
the function $H(z)$ must have one or more discontinuities. In the simplest configuration, a single discontinuity is introduced at the cigar tip, $z=0$, and the function $H(z)$ is taken to be odd in $z$. In the UV, $|z|\to\infty$, it must vanish for the configuration to have finite energy density. The baryon chemical potential $\mu_B$ is encoded in the UV value of the abelian field, and the baryon {number} density $n_B$ is related to the discontinuity in $H(z)$ as:
\begin{align}
&\mu_B =\frac{N_c}{2} \widehat{a}_0 (z\rightarrow\pm\infty),\\
&n_B = \frac{1}{8\pi^2}\left[H(z\rightarrow0^+)-H(z\rightarrow0^-)\right].
\end{align}
The pressure $P$ of the system can be calculated from the standard holographic dictionary by evaluating the on-shell action. The energy density $\mathcal{E}$ is then derived from the thermodynamic relation:
\begin{equation}
    P=-\mathcal{E}+\mu_B n_B.
\end{equation}
Alternatively, the energy density can also be computed via the matter stress-energy tensor.

The practical appeal of this construction is clear: the bulk PDEs reduce to ODEs and one can scan large regions of parameter space with standard numerical methods. At the same time, the approximation has certain unresolved conceptual issues. 
In the WSS model it leads to discontinuities and to an ambiguity in the Chern-Simons term, which can be cured only by supplementing the action with a boundary term in order to recover the correct conserved charges \cite{Bartolini:2023eam}. Even if a homogeneous structure, as described by Eq.~\eqref{eq:homansatz} together with its discontinuities, were to emerge from a multi-soliton configuration in the high-density limit, it is unclear at what densities the approximation would be appropriate. Densities of the order of the saturation density $n_S\approx0.16\rm \; fm^{-3}$, which are expected in realistic neutron stars, may or may not be large enough for the homogeneous approximation to be suitable.

An additional drawback of the homogeneous ansatz is its insensitivity to the quark mass, when introduced via the term \eqref{eq:AK}. Since $A_z=0$ for the smeared holographic nuclear matter, the equations of motion for arbitrary quark masses are still solved by the same homogeneous configuration. All equilibrium observables will then be independent of the quark mass, contrary to expectations from nuclear EFT.

The phenomenological limitations are equally serious: when the model parameters are fixed to mesonic observables (or, as in V-QCD, from input by lattice-QCD), the homogeneous construction 
is generally unable to reproduce
basic nuclear-matter observables. The details are model dependent, but the pattern is robust: the saturation density comes out much too large in WSS and hard-wall models, much too small in V-QCD, and the symmetry energy is typically too large in both models \cite{Kovensky:2021ddl,Kovensky:2021kzl,Kovensky:2021wzu,Bartolini:2022gdf,Bartolini:2022rkl,Jokela:2018ers,Jarvinen:2021jbd,Bartolini:2025sag}. 
As a result, even when the pressure-energy curve can be made compatible with neutron-star bounds at high density, 
a continuous matching of the energy, pressure and chemical potential to the low-density phenomenological EOS band obtained from chiral EFT is often not possible. This has also recently been observed in D3-D7 models in Ref.~\cite{BitaghsirFadafan:2025rpm}, where baryons are introduced as D5 branes\footnote{These models also predict a quark phase at densities possibly relevant in the centers of the heaviest neutron stars.}. The homogeneous approximation does not require discontinuities, but is still unable to reproduce basic properties of (symmetric) nuclear matter at saturation.

\begin{figure}
    \centering
    \includegraphics[width=0.49\linewidth]{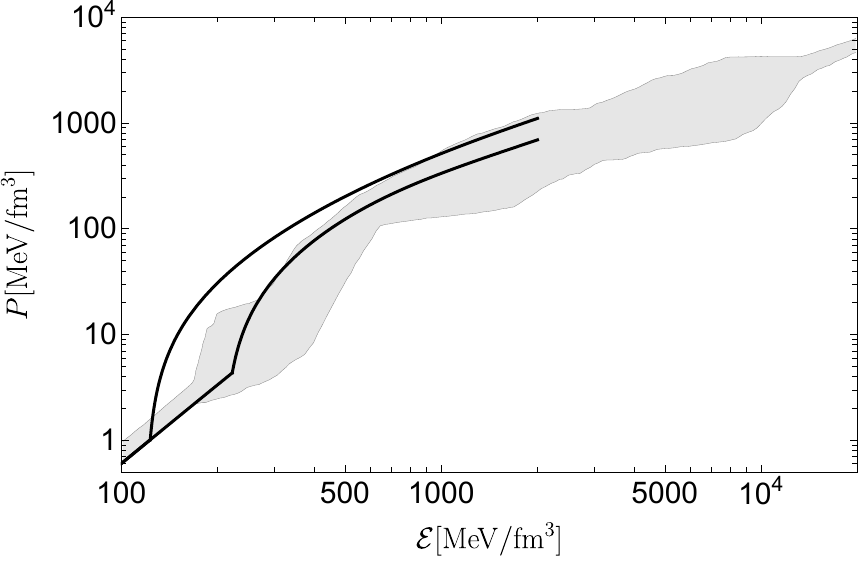}
    \includegraphics[width=0.49\linewidth]{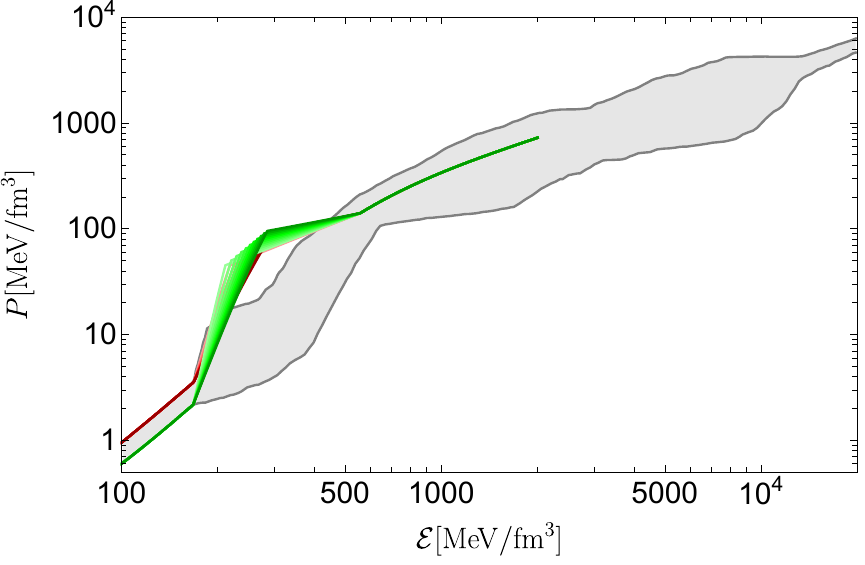}
    \caption{\textbf{Left:} Hybrid EOS from the homogeneous ansatz for $\lambda=16.63$ (stiffer; upper curve) and for $\lambda=0.5$ (softer; lower curve). \textbf{Right:} Various hybrid EOSs with a piecewise polytropic interpolation between the low-density phenomenological EOS and the high-density holographic one. Every possible construction violates the bound from neutron stars observations and results in an EOS that is far too stiff.}
    \label{fig:HAEOS_Polytwopoints}
\end{figure}

To make this tension explicit, we construct a hybrid EOS using the Hebeler-Lattimer-Pethick-Schwenk (HLPS) construction \cite{Hebeler:2013nza} for the low-density regime and the homogeneous holographic EOS of the WSS model for the high-density region. We require continuity of pressure, energy density, and chemical potential when matching, while continuity of baryon number density then follows automatically. We choose a polytropic index $\Gamma=2.5$ for the HLPS construction beyond $n_B=n_1=1.1\,n_S$, and we choose the softest variant for the region at $n_B<1.1 \,n_S$. 
Starting from the standard mesonic value $\lambda=16.63$ and enforcing the continuity conditions to fix $M_{\rm KK}$ and the transition density $n_t$, we obtain $M_{\rm KK}\simeq 630.5\, \rm MeV$ and $n_t\simeq 0.8 \,n_S$. The transition therefore occurs below saturation, precisely where the homogeneous approximation is not trustworthy. Additionally, as shown in the left panel of Fig.~\ref{fig:HAEOS_Polytwopoints}, the resulting EOS $P(\mathcal{E})$ becomes unrealistically stiff.

Although decreasing the 't Hooft coupling $\lambda$ softens the EOS and pushes $n_t$ to higher values, new problems are introduced.
Continuous matching with the low-density construction would require very small values of the 't Hooft coupling $\lambda \lesssim 0.5$ and exceedingly large values of the mass scale $M_{\rm KK}\sim 4\,\rm GeV$. These values are not only extreme when compared to the usual mesonic sector but also problematic from the top-down perspective (assuming $\lambda\gg1$), as stringy corrections would likely become non-negligible. In the left panel of Fig.~\ref{fig:HAEOS_Polytwopoints} we show the construction described with $(\lambda,M_{\rm KK})=(16.63,630.5\,\mathrm{MeV})$ and $(0.5,4120\,\mathrm{MeV})$.

In the right panel of the same figure, we show an attempt to push the transition density to $n_t=3\, n_S$. Below $3n_S$, two polytropic sections with free polytropic indices are used. Requiring continuity of $P,\mathcal{E},\mu_B$ in the matching procedure results in too stiff equations of state, which give rise to too heavy and too large neutron stars and violate constraints from nuclear physics near saturation density.

Several workarounds have been explored in the literature. In the WSS model, one possibility is to require continuity of pressure and number density while relaxing continuity of the energy density, thereby introducing a first-order transition slightly above saturation \cite{Bartolini:2023wis}. Although such a first-order transition is a logical possibility, the description of dense nuclear matter as soliton crystals, as presented in Sec.~\ref{sec: nuclear matter}, suggests a much smoother behavior. Another approach within the WSS model describes the low-density region holographically through a mixed phase of homogeneous matter and leptons \cite{Kovensky:2021kzl}. The surface tension of the domain wall is taken from phenomenology  but saturation densities remain significantly above the physical values.
In hard-wall models, the homogeneous construction has been used only for the dense core of the star \cite{Bartolini:2022rkl}, without involving the construction of a hybrid EOS. In V-QCD, continuous matching can only be facilitated by introducing an additional parameter that rescales the action and is fitted in the baryonic phase \cite{Jokela:2018ers,Jarvinen:2021jbd,Jarvinen:2023jbr,Bartolini:2025sag}. These studies show that the homogeneous ansatz remains a useful exploratory tool, but also that obtaining phenomenologically viable baryonic physics typically requires compromises, either in the form of ad-hoc first-order transitions, additional free parameters, or a limited description of saturation properties.

These issues of the homogeneous ansatz motivate a more localized holographic description of nucleons.  As shown in the next section, crystals of nucleons with interactions extracted from the WSS model are in much better agreement with low-energy phenomenology. Additionally, it is much simpler to judge the validity of this description, in contrast to the homogeneous method.

\section{Holographic soliton crystals}\label{sec: nuclear matter}

Instead of employing the homogeneous ansatz, we 
will construct a configuration of infinitely extended nuclear matter that preserves a connection to the solitonic single-baryon description. 
In order to describe their interactions,
we rely on the ideas presented in Refs.~\cite{Baldino:2017mqq, Baldino:2021uie}
to obtain a controllable approximation to the nucleon-nucleon potential, which
we then employ
to compute the energy density and free energy density of an infinite lattice of interacting baryons.

In Ref.~\cite{BitaghsirFadafan:2018uzs}, where also soliton crystals were
considered, the interactions were
approximated by starting from the ADHM construction \cite{Atiyah:1978ri}
of two-instanton
solutions but then employing spatial averages of their respective field
strengths squared such that a spatially homogeneous description arises. 
Moreover, a uniform orientation was assumed.

Following Refs.~\cite{Baldino:2017mqq, Baldino:2021uie}, 
we consider two baryons, whose centers are located at $(Z=0)$ and  are separated by $\vec{r}$ in $\mathbb{R}^3$, with $|\vec{r}|\sim\mathcal{O}(1)$ in units of $M_{\rm KK}^{-1}$. Recall that the single instanton size $\rho$ is of order $\rho\sim\mathcal{O}(\lambda^{-{1}/{2}})$, so that in the large $\lambda$ limit $|\vec{r}|\gg\rho$. In this scenario, the full gauge field is approximately given by $\mathcal{A}=\mathcal{A}^p+\mathcal{A}^q$, where $\mathcal{A}^p$ is the field of the soliton centered at $\vec{X}_1$, and $\mathcal{A}^q$ is the field of the one sitting at $\vec{X}_2$, with $\vec{r}=\vec{X}_1-\vec{X}_2$. We  then divide the space $\mathbb{R}^3$ into three regions $P$, $Q$ and $\mathbb R^3-P-Q$: the $P$ and $Q$ regions are given by balls of radius $\sim \lambda^{-1/2}$ centered around the two solitons  respectively. In the $P$ region, we may approximate $\mathcal{A}^p$ by the BPST instanton solution, and similarly for $\mathcal A^q$ in region $Q$. We assume that in the region $\mathbb R^3-P$, the instanton solution $\mathcal{A}^p$ is well approximated by the linearized far-field expressions \eqref{eq:farfield}, \eqref{eq:farfield2} and \eqref{eq:farfield3}, and analogously for $\mathcal{A}^q$ in $\mathbb R^3-Q$.

\begin{figure}
    \centering
    \includegraphics[width=0.49\linewidth]{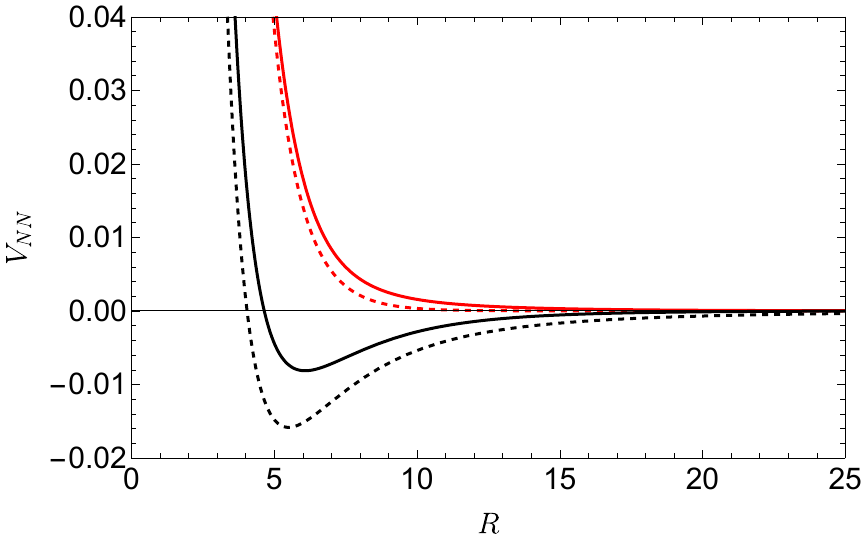}
    \includegraphics[width=0.49\linewidth]{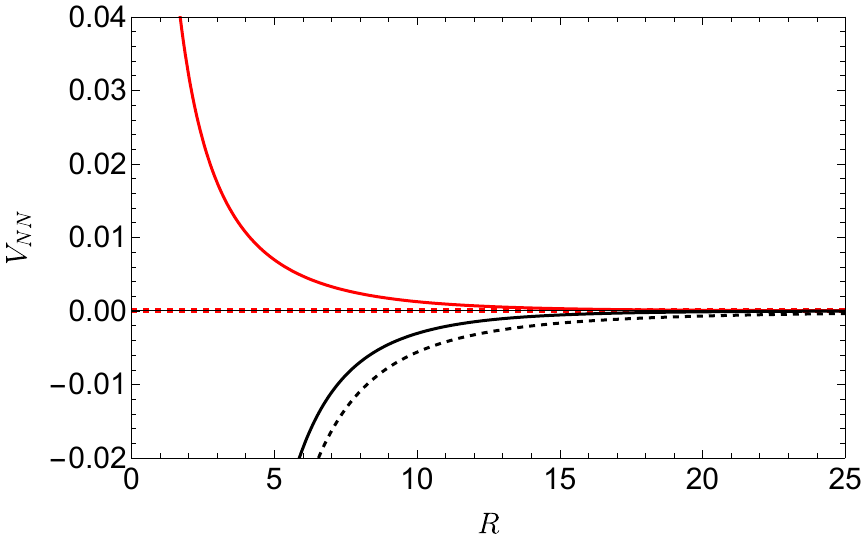}
    \caption{\textbf{Left:} The nucleon-nucleon potential \eqref{eq:dinuclpotential} in dimensionless units. Black lines indicate the maximally attractive channel selected by setting $\alpha=\pi,\,\hat{u}\perp \vec{R}$ in Eq.~\eqref{eq:mmatrixaxisrotation}, while red lines correspond to the repulsive channel selected by $\alpha=0$ and corresponding to $M_{ij}=\delta_{ij}$. Solid lines show results for physical pion mass $m_\pi=135\, \rm MeV$, dashed lines correspond to the chiral limit. \textbf{Right:} The pion-mediated contribution to Eq.~\eqref{eq:dinuclpotential}. The contribution vanishes in the chiral limit of the repulsive channel, and is screened by the pion mass in the maximally attractive one. All curves are calculated at $\lambda=18.26$.}
    \label{fig:Potentials}
\end{figure}

The two individual solitons can be given arbitrary orientations by writing
\begin{align}
    \mathcal{A}^p= B \mathcal{A}^I(\vec{x}-\vec{X}_1)B^\dagger,\qquad  \mathcal{A}^q= C \mathcal{A}^I(\vec{x}-\vec{X}_2)C^\dagger,
\end{align}
where $\mathcal{A}^I$ indicates the gauge field of a single soliton with standard orientation and $C,B \in SU(2)$ are constants.

Using the equations of motion and integrating by parts, the energy of any static configuration is given by
\begin{align}
   \nonumber E=&\frac{\lambda N_c}{216\pi^3}\int dx\,dz\;\left[\frac{1}{2}h(z)\text{tr}\left(F_{ij}^2\right)+k(z)F_{iz}^2-\frac{1}{2}\widehat{A}_0\left[h(z)\partial_i\partial_i+\partial_z \left(k(z)\partial_z\right)\right]\widehat{A}_0\right] +\\
    &+a\int d^3x\,dz\;\text{tr}\left(\int_{-\infty}^{z} dz'\, A_z(x,z)A_z(x,z')\right).
\end{align}
Inserting the ansatz $\mathcal{A}=\mathcal{A}^p+\mathcal{A}^q$, the integral can be divided into two self-energies (involving only either $\mathcal A^p$ or $\mathcal{A}^q$) and an interaction term $V(\vec{r},B,C)$. The self-interactions will evaluate to the mass of the two isolated baryons, so that the energy is given by $E= 2M_B + \hat V(\vec{r},B,C)$.
Following 
Refs.~\cite{Baldino:2017mqq,Baldino:2021uie}, we obtain the interaction potential\footnote{This corrects eq.~(3.6) of Ref.~\cite{Baldino:2021uie}, where
$P^\pi_{ij}$ was erroneously identified with $P_{ij}$; the corrected expression, $P^\pi_{ij}$, is consistent with the result of Ref.~\cite{Gudnason:2020arj}.
}
\begin{align}
\label{eq:dinuclpotential}
\hat V(r,B, C) =\frac{27\pi^2 N_c}{2 \lambda}&\Bigg[\left( \sum_{n=1}^{\infty}\frac{1}{c_{2n-1}}\frac{e^{-k_{2n-1}r}}{r}+\sum_{n=1}^{\infty}\zeta\frac{1}{b_n}M_{ij}(B^\dagger C)P_{ij}(\vec r,k_n)\frac{e^{-k_n r}}{r^3}\right) \nonumber\\
&\quad \mathop-\zeta \frac{1}{\pi}M_{ij}(B^\dagger C)P^{\pi}_{ij}(\vec r,k_0)\frac{e^{-k_0 r}}{r^3}\Bigg],
\end{align}
where 
\begin{align}
    &M_{ij}(G)=\frac{1}{2}\text{tr}\big(\sigma_i G \sigma_j G^\dagger \big),\\
    &P^\pi_{ij}(\vec r,k)=\delta_{ij}\left(rk+1\right) -\frac{r_i r_j}{r^2}\left((rk)^2+3rk+3\right),\\
    &P_{ij}(\vec r,k)=\delta_{ij}(rk)^2 + P^\pi_{ij}(\vec r,k),
\end{align}
and 
\begin{equation}
\zeta=\frac{\lambda^2\rho^4}{3^6\pi^2}=\frac65\left(1-\frac{36 \mathcal{I}}{\sqrt{\pi}}\left(\frac{6}{5}\right)^{\frac{1}{4}}\frac{m}{N_c}\right)^4, \qquad 
 b_n= \begin{cases}
     -d_n &\qquad n=\rm even,\\
     c_n  &\qquad n=\rm odd,
 \end{cases}
\end{equation}
with $c_n,d_n$ defined in Eq.~\eqref{eq:modeeq}
and $\mathcal{I}$ below in Eq.~\eqref{calI}.
The sum over $i,j$ in Eq.~\eqref{eq:dinuclpotential} is implicit and the potential $\hat{V}(r,B,C)\equiv V(r,B^\dagger C)$ only depends on the combination $B^\dagger C$ of the individual orientations. In Fig.~\ref{fig:Potentials} we plot the potential \eqref{eq:dinuclpotential} for the maximally attractive (``combed Skyrmions" \cite{Kim:2008iy}) and repulsive channels (``defensive Skyrmions" \cite{Kim:2008iy}) for both massless and massive pions.

The nucleon-nucleon potential can thus be naturally interpreted in terms of massive meson exchange.
The first sum in Eq.~\eqref{eq:dinuclpotential} is a repulsive monopole interaction mediated by the tower of isoscalar $\omega$ vector mesons.
The second sum
describes a dipole interaction to which isovector vector and axial-vector mesons (odd and even indices, respectively) contribute with opposite sign. The last term is the contribution from the pseudo-Goldstone pion multiplet\footnote{In the chiral limit (which is not taken in this paper), this term would induce long-range interactions.}. 
The potential \eqref{eq:dinuclpotential} supports a first-order baryonic onset, as shown in Ref.~\cite{Baldino:2021uie}, while it is known from Ref.~\cite{Bergman:2007wp} that pointlike instantons produce a second-order transition to baryonic matter: the pointlike instantons result is consistently recovered in our framework by taking the limit $\rho\rightarrow0$, which causes the dipole terms in Eq.~\eqref{eq:dinuclpotential} to vanish, leaving only the monopole term which originates from the abelian field. The monopole term is universally repulsive and as such would lead to a second-order baryonic onset.

We use this potential to build an infinite lattice with a face-centered cubic (FCC) unit cell, which provides a maximally close packing of spheres in three dimensions. {A comparison to the simple cubic (SC) lattice considered originally by Klebanov \cite{Klebanov:1985qi}, as well as a discussion of other lattice structures can be found in appendix \ref{app:latticegeometry}}. Nearest neighbors, whose distance to each other is denoted by $R$, are chosen to have their relative $SU(2)$ orientations in the maximally attractive channel. The length of the unit cell is given by $\sqrt 2 R$. In the FCC geometry, the $12$ nearest-neighbor separation vectors can be written as
\begin{equation}
    \vec{r}=\frac{R}{\sqrt{2}}(\pm 1,\pm 1,0)\,,\qquad \text{and permutations,}
\end{equation}
with independent sign choices.

\begin{figure}
    \centering
    \includegraphics[width=0.40\linewidth]{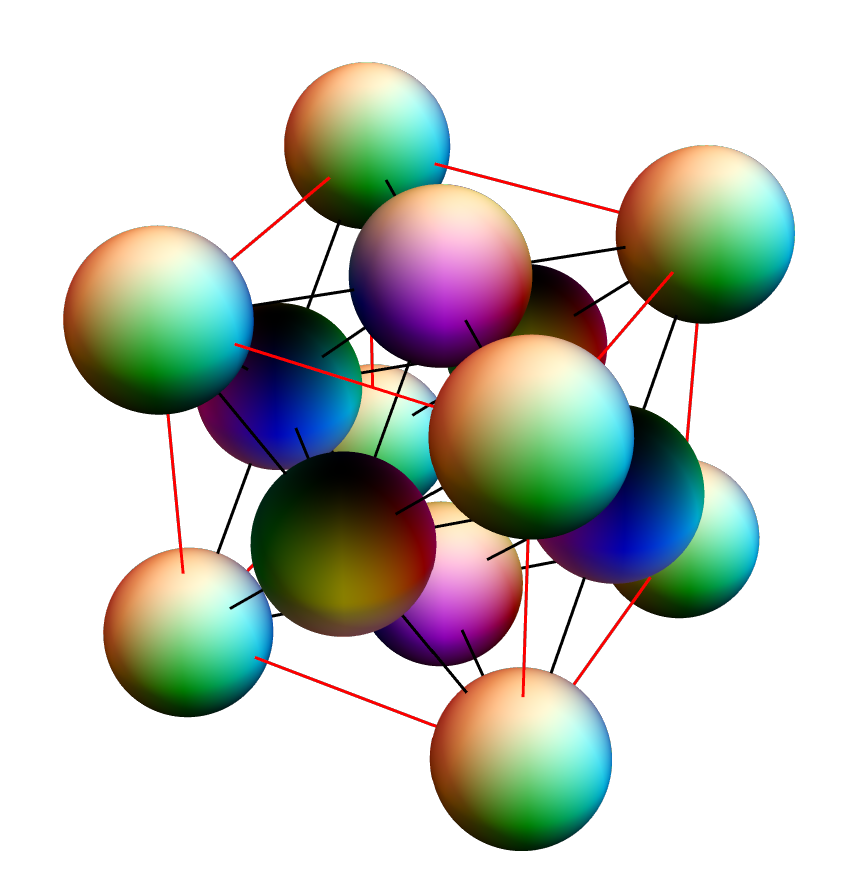}
    \includegraphics[width=0.58\linewidth]{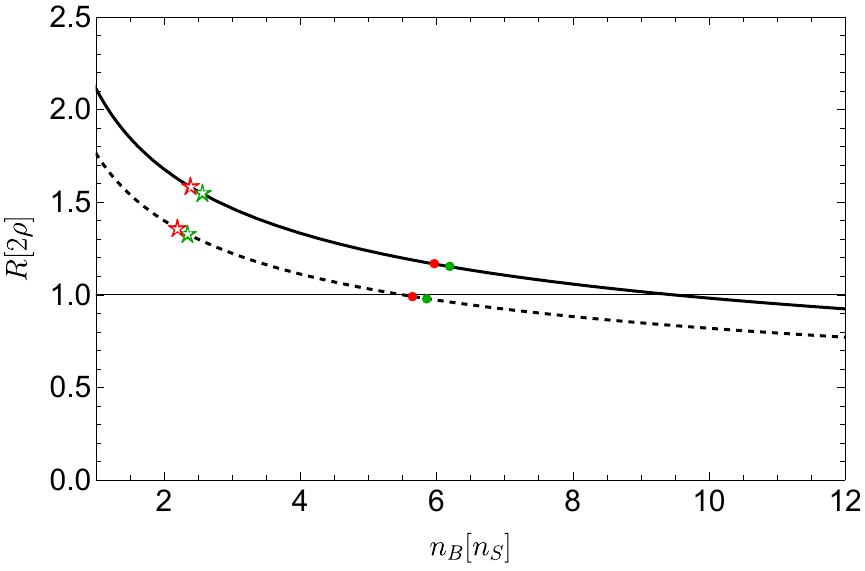}
    \caption{\textbf{Left:} FCC unit cell and $SU(2)$ orientation assignment used for the solitonic baryons in the crystal. Each sphere represents the instanton core of a baryon, and the colors encode the relative $SU(2)$ orientations: white is $\i\sigma^3$, black is $-\i\sigma^3$, red is $\i\sigma^1$, yellow is $\frac{\i}{2}(\sigma^1+\sqrt{3}\sigma^2)$, green is $\frac{\i}{2}(-\sigma^1+\sqrt{3}\sigma^2)$, cyan is $-\i\sigma^1$, blue is $\frac{\i}{2}(-\sigma^1-\sqrt{3}\sigma^2)$ and magenta is $\frac{\i}{2}(\sigma^1-\sqrt{3}\sigma^2)$. The red lines delimit the unit cell, within which are contained $B=4$ baryons (each corner contributing with $1/8$ and each face contributing with $1/2$ of a baryon). \textbf{Right:} Separation between nearest neighbors in units of $2\rho$, as a function of density (in units of saturation density $n_S$). The solid (dashed) line corresponds to the BPST (non-BPST) choice for the parameters $(\gamma,\eta)$. For both fit choices, the separation between the soliton centers remains larger than twice their size around the saturation. The red and green dots represent the highest density reached at the center of stable stars, depending on the symmetry energy parameters $S_0$ and $L$ (see section 6), while the red and green stars represent the values for the highest density reached in a NS of $1.4 M_\odot$.
    }
    \label{fig:Lattice}
\end{figure}

Parametrizing $SU(2)$ elements by
\begin{align}
    G=e^{\i\alpha^a t^a}, \quad t^a=\sigma^a/2,
\end{align}
and defining $\alpha= \sqrt{\alpha^a\alpha^a}$ and $\hat u^a =\alpha^a/\alpha$, the attractive channel corresponds to a relative $SU(2)$ rotation by $\alpha=\pi$. It is useful to write $M_{ij}$ in the axis-angle notation,
\begin{equation}\label{eq:mmatrixaxisrotation}
    M_{ij}(\hat{u},\alpha) = \delta_{ij}\cos\alpha +\left(1-\cos\alpha\right)\hat{u}_i\hat{u}_j+\epsilon_{ijk}\hat{u}_k\sin\alpha.
\end{equation}
For the nearest neighbors of the FCC crystal, the relative iso-orientation $B^\dagger C$ is taken to be a rotation by $\alpha=\pi$ around the coordinate axis corresponding to the vanishing entry in $\vec{r}$. Explicitly, we choose $B^\dagger C=\pm\i\sigma^k$ with $k$ determined by the zero component of $\vec{r}$:
\begin{itemize}
\item $B^\dagger C=\pm\i\sigma^3$ for $\vec{r}=\frac{R}{\sqrt{2}}(\pm 1,\pm 1,0)$ (rotation axis $\hat{u}=\hat{x}_3$),
\item $B^\dagger C=\pm\i\sigma^1$ for $\vec{r}=\frac{R}{\sqrt{2}}(0,\pm 1,\pm 1)$ (rotation axis $\hat{u}=\hat{x}_1$),
\item $B^\dagger C=\pm\i\sigma^2$ for $\vec{r}=\frac{R}{\sqrt{2}}(\pm 1,0,\pm 1)$ (rotation axis $\hat{u}=\hat{x}_2$),
\end{itemize}
see Fig.~\ref{fig:Lattice} for a visual representation. 
The FCC unit cell contains baryon number $B=4$.
The energy per baryon (which we label $E_p$) is given by
\begin{equation}\label{eq:baryonenergy}
    E_p = \frac{1}{2}\sum_{q\neq p} V\left(\vec{r}_p -\vec{r}_q,B_p^\dagger C_q\right) + M_B,
\end{equation}
where we include a factor of one half in the interaction term to avoid double counting, and $M_B$ is the nucleon mass obtained from the WSS model. We use a mass formula motivated by the instantonic description of the baryon, which also captures corrections to the approximate expression of Ref.~\cite{Hata:2007mb}. In units of $M_{\rm KK}$ the mass reads
\begin{equation}\label{eq:baryonmass}
    M_B= \left(\frac{\lambda N_c}{27 \pi} + \sqrt{\frac{2}{15}}N_c\right)\gamma + \frac{16 \lambda^{\frac{3}{2}}}{3^{\frac{9}{2}}\pi^2}\rho_{\rm cl}^3\eta^3\mathcal{I}m.
\end{equation}
The two terms in Eq.~\eqref{eq:baryonmass} originate respectively from the static soliton energy and the mass shift due to the nonvanishing quark mass \cite{Hashimoto:2009hj}, and $\mathcal{I}$ is given by the integral 
\begin{equation}\label{calI}
    \mathcal{I} = \int^\infty_0 dy y^2 \left( 1-\cos \left[\pi \left(1-\frac{1}{\sqrt{1+y^{-2}}}\right)\right] \right)\approx 1.104,
\end{equation}
obtained from the radial integral of the holonomy of $A_z$ on the baryon background, after rescaling the radial coordinate with $\rho$ (which gives the overall $\rho^3$ dependence). The factors $\gamma$ and $\eta$   are not present in the BPST-instanton-based calculation of Refs.~\cite{Hata:2007mb,Hashimoto:2009hj}. They are introduced here to account for curvature corrections to the formulas of Ref.~\cite{Hata:2007mb}, which were computed numerically in Ref.~\cite{Hori:2023fxq} for the WSS model fitted to $M_\rho$ and $f_\pi$, albeit with massless pions.

The extrapolation of the flat-space large-$\lambda $ result of Ref.~\cite{Hata:2007mb} to $\lambda=16.63$ and $N_c=3$, gives $M_I^{\rm HSSY}=1.68$ and $\rho_I^{\rm HSSY}=2.36$ (in units of powers of $M_{\rm KK}$).

Ref.~\cite{Hori:2023fxq} considered two different situations. In the first case, the same BPST instanton, supplemented by an abelian field sourced from the Chern-Simons term, as Eq.~\eqref{eq:BPSTplusabelian} was used to compute the energy numerically, yielding $M_I^{\rm HSK,1}=1.35,\;\rho_I^{\rm HSK,1}=1.8$. In the second case, the solitonic field configuration, as well as its energy, were computed numerically resulting in $M_I^{\rm HSK,2}=1.25,\;\rho_I^{\rm HSK,2}=2.4$. The true solution of the equations of motion thus predicts lighter and spatially bigger baryons.
We then define
\begin{align}
    &\gamma \equiv \left( \frac{M_I^{\rm HSK}}{M_I^{\rm HSSY}} \right)_{\lambda=16.63},\qquad \eta\equiv \left( \frac{\rho^{\rm HSK}}{\rho^{\rm HSSY}} \right)_{\lambda=16.63},
\end{align}
and use the following two parameter sets:
\begin{align}
\label{eq:ch1}
    &\text{set 1 (BPST):}\phantom{\text{non-}}\quad M_I^{\rm HSK}=1.35,\;\rho_I^{\rm HSK}=1.8\quad\Rightarrow\quad\gamma\simeq 0.80,\;\eta\simeq 0.76,\\
    \label{eq:ch2}
    &\text{set 2 (non-BPST):}\quad M_I^{\rm HSK}=1.25,\;\rho_I^{\rm HSK}=2.4\quad\Rightarrow\quad\gamma\simeq 0.74,\;\eta\simeq 1.02.
\end{align}
The rescaling factors $\eta$ and $\gamma$ are then assumed to be 
approximately independent of $\lambda$. Any appearance of the instanton size $\rho$ is also  rescaled as $\rho$ as $\rho\rightarrow\eta \rho$, including the dipole terms in the potential \eqref{eq:dinuclpotential}.

To make the lattice sum numerically tractable, we introduce a cutoff by grouping lattice sites into ``shells'' of fixed distance from the reference baryon. It is convenient to parameterize FCC lattice vectors as
\begin{equation}
    \vec{r}=\frac{R}{\sqrt{2}}(u,v,w),\qquad u,v,w\in\mathbb{Z},\qquad u+v+w\in 2\mathbb{Z},
\end{equation}
so that the squared distance from the reference baryon is
\begin{equation}
    |\vec{r}|^2=\frac{R^2}{2}\left(u^2+v^2+w^2\right)\equiv n R^2,\qquad n\in\mathbb{N}.
\end{equation}
A shell is then labeled by the integer $n$ (i.e. $|\vec{r}|=\sqrt{n}\,R$). The contribution of shell $n$ to the interaction energy is obtained by summing $V\big(|\vec{r}|,B^\dagger C\big)$ over all integer triples $(u,v,w)$ satisfying $u^2+v^2+w^2=2n$, multiplied by the multiplicity of vectors related by sign flips and by permutations that leave the class $(|u|,|v|,|w|)$ invariant.
For the iso-orientation assignment of the FCC crystal, shells with even $n$ have $B^\dagger C=\mathds{1}$, whereas for odd $n$ one has a relative rotation by $\pi$ about a coordinate axis determined by the unique even entry in $(u,v,w)$ (equivalently: by the vanishing entry in the nearest-neighbor case). More details on the $SU(2)$ orientations in the FCC lattice can be found in appendix \ref{app:FCC-shell-counting}.
In practice we truncate the shell sum at a maximal shell index, $n\leq n_{\max}=30$.

The energy associated to all the lattice is then simply obtained by summing over each baryon $p$, which will simply give an overall factor of the total baryon number $B$ as a consequence of the symmetry of the lattice: 
\begin{equation}\label{eq:latticeenergy}
    E= \sum_{p} E_p =  B E_p = \frac{\sqrt{2}\,V_L}{R^3} E_p.
\end{equation}
Since we are describing an infinitely extended system, it is useful to work with intensive quantities, so we define the energy density of the system as the total energy divided by the volume of the lattice $V_L$. The FCC unit-cell volume is $V_{\rm cell}=2\sqrt{2}\,R^3$ and it contains baryon number $B=4$, hence the number density per unit cell is
\begin{equation}
    n_B=\frac{4}{V_{\rm cell}}=\frac{\sqrt{2}}{R^3}.
\end{equation}
We then express the energy density in terms of either $R$ or $n_B$:
\begin{equation}
\label{eq:energydensitysymm}
    \mathcal{E}=\frac{E}{V_L} = \frac{\sqrt{2}}{R^3} E_p(R) = n_B E_p (R(n_B)), \qquad R(n_B) = \left(\frac{\sqrt{2}}{n_B}\right)^{\frac{1}{3}}.
\end{equation}
From the energy density, we can then compute the chemical potential $\mu_B$ and the free energy density $\Omega$:
\begin{equation}
    \mu_B = \frac{\partial \mathcal{E}}{\partial n_B},\qquad \Omega = -P = \mathcal{E} - \mu_B n_B.
\end{equation}

\section{Parameter fits and properties of symmetric nuclear matter}
\label{sec:symmetricfit}

The FCC construction of the previous section provides an isospin-symmetric nuclear matter equation of state. For each of the two choices \eqref{eq:ch1} and \eqref{eq:ch2} of $(\gamma,\eta)$, we fit the free parameters $\lambda,M_{\rm KK}$ to the onset baryon chemical potential $\mu=922.7$ MeV and such that the onset number density is given by $n_S=0.16 \;\rm fm^{-3}$.

For the BPST-derived choice of $(\gamma,\eta)$, one obtains
\begin{equation}
    \lambda=18.26,\qquad M_{\rm KK}=661.5\,\mathrm{MeV}.
\end{equation}
The value for the 't Hooft coupling $\lambda$ is close to the standard mesonic calibration \cite{Sakai:2004cn}, $\lambda=16.63$, while $M_{\rm KK}$ is somewhat lower but still in the right ballpark. 
These parameters in fact correspond to a simultaneous reduction of both $f_\pi$ and $k_1\equiv M_{\rho,\omega}$, to 68 and 540 MeV, respectively, which could be interpreted as an in-medium reduction in line with the expectations
from Brown-Rho scaling \cite{Brown:1991kk}, namely $f_\pi^\mathrm{med}/f_\pi^\mathrm{vac}\sim M_{\rho,\omega}^\mathrm{med}/M_{\rho,\omega}^\mathrm{vac}<1$.

With this fit, the (single) baryon mass from Eq.~\eqref{eq:baryonmass} is 
\begin{equation}
    M_B=942.0\,\mathrm{MeV},
\end{equation}
and the binding energy per nucleon at saturation is
\begin{equation}
    \frac{E}{A}-M_B=-19.3\,\mathrm{MeV},
\end{equation}
which both are reasonably close to phenomenology (for isospin symmetric nuclear matter) \cite{Stone:2024inj}.

For the incompressibility at saturation we obtain
\begin{equation}
    K(n_S)=374.9\,\mathrm{MeV}.
\end{equation}
This is slightly larger than the commonly quoted upper bound ($\sim 300\,\mathrm{MeV}$), but the order of magnitude is correct, representing a major improvement over the homogeneous ansatz, which overestimates $K(n_S)$ by about an order of magnitude even with multi-jump configurations \cite{Ecker:2025sjb}.

For the non-BPST choice of $(\gamma,\eta)$, the fit parameters are
\begin{equation}
    \lambda=45.54,\qquad M_{\rm KK}=457.2\,\mathrm{MeV},
\end{equation}
also implying an (in-medium) reduction of both $f_\pi$ and $M_{\rho,\omega}$, now
to 74 and 373 MeV, respectively.
The baryonic observables 
\begin{align}
    M_B&=958.5\,\mathrm{MeV},\\
    \frac{E}{A}-M_B&=-35.8\,\mathrm{MeV},\\
    K(n_S)&=517.6\,\mathrm{MeV},
\end{align}
agree
less well with real-world nuclear-matter properties than those obtained with the BPST choice of $(\gamma,\eta)$. It should also be noted that this fit requires $\lambda=45.54$, which is considerably larger than the standard mesonic-fit value, $\lambda=16.63$.
The values for $\eta$ in the two different fits differ by roughly $30 \%$, but the interaction \eqref{eq:dinuclpotential} contains dipole terms proportional to $\rho^4$, leading to the significant changes in $\lambda,M_{\rm KK}$. A potential reason could be that the tails of the instanton, which end up determining the interaction potential, are derived from matching to BPST instantons, while baryon mass and size are phenomenologically modified to reflect the numerical non-BPST solutions determined in Ref.~\cite{Hori:2023fxq}. Whether this is responsible for the phenomenologically less viable fit could be ascertained from numerical calculations of the full multi-soliton configuration. While this is not pursued in the current exploratory paper, it remains an interesting question for future studies. 

Despite the differences between the BPST and non-BPST fits, the associated equations of state (EOS) can nonetheless both be matched (almost) continuously to the low-energy chiral EFT EOS, after inclusion of isospin-breaking effects, as shown in Sec.~\ref{sec:betaeq}. There is a clear improvement with regards to the homogeneous ansatz, as shown in App.~\ref{app:equalfooting}.

\begin{figure}
    \centering
    \includegraphics[width=0.497\linewidth]{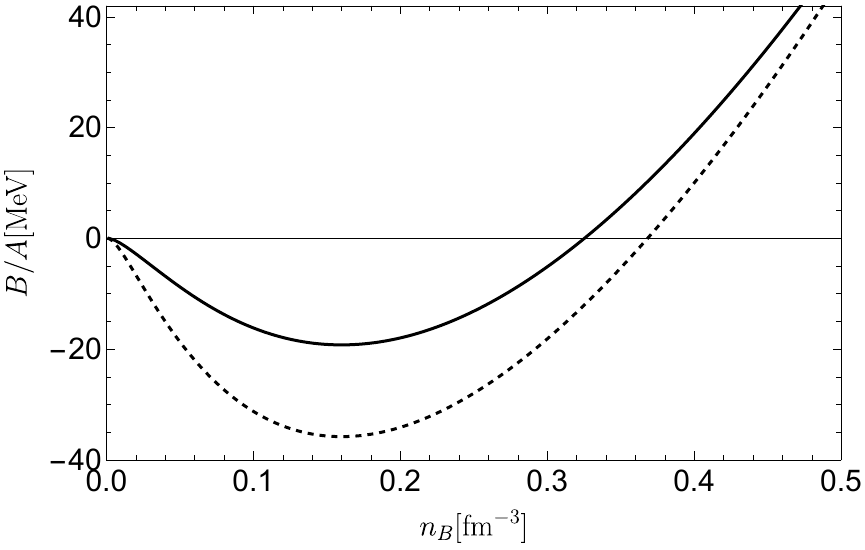}
    \includegraphics[width=0.483\linewidth]{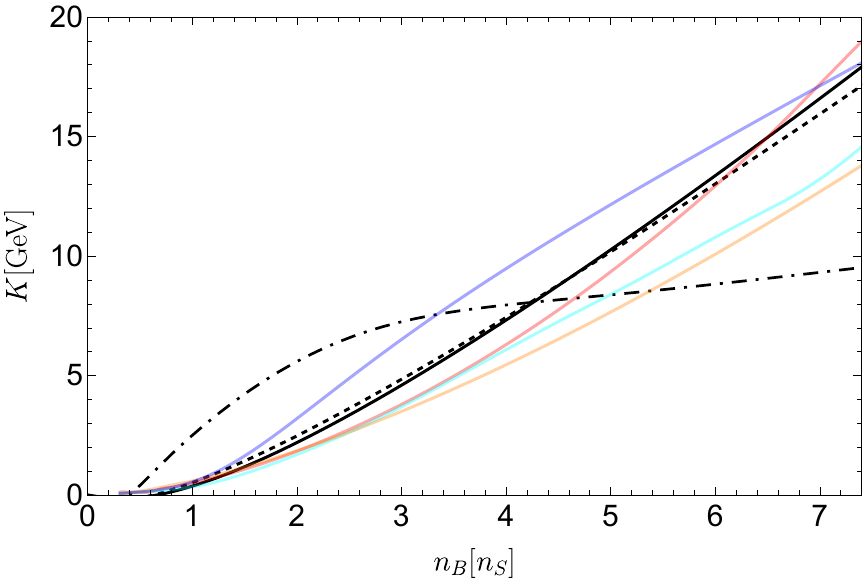}
    \caption{\textbf{Left:} Binding energy per nucleon for the calibrated symmetric-matter EOS. \textbf{Right:} Incompressibility as a function of baryon density in units of saturation density. The black dot-dashed line corresponds to the single-layer homogeneous configuration calibrated at $\lambda=5.79$ and $M_{\rm KK}=1119.33\,\mathrm{MeV}$, obtained by fitting saturation density and the chemical potential at onset (note that in Ref.~\cite{Ecker:2025sjb} the values reported are $\lambda=7.09$, $M_{KK}=1000$: the discrepancy is due to our choice of $n_S=0.16\,\rm fm^{-3}$, as opposed to $n_S=0.15\,\rm fm^{-3}$ used by the authors of Ref.~\cite{Ecker:2025sjb}). The light colored lines show the incompressibility computed in Ref.~\cite{Perego:2021mkd} for a selection of non-holographic models of nuclear matter: as in Fig.~10 of Ref.~\cite{Ecker:2025sjb} we include BL (red), DD2 (blue), LS220 (orange) and SFHo (light blue). References for the individual models can be found in Ref.~\cite{Perego:2021mkd}. For all panels, the solid (dashed) black line corresponds to the BPST (non-BPST) choice of parameters $(\gamma,\eta)$.}
    \label{fig:Binding_K}
\end{figure}

Another clear improvement is found in the density dependence of the incompressibility: as can be seen in the right panel of Fig.~\ref{fig:Binding_K}, the qualitative behavior of $K(n_B)$ is in agreement with that derived from selected non holographic models, in stark contrast with that computed from the homogeneous ansatz, which shows a steeper increase around saturation density and then a milder density dependence above roughly $3n_S$.

A further qualitative point concerns the density dependence of the incompressibility, presented in Fig.~\ref{fig:Binding_K}. In the FCC-crystal construction, the holographic incompressibility exhibits a density trend that is qualitatively similar to that found in non-holographic nuclear-matter modeling \cite{Perego:2021mkd}. This is in contrast with the milder high-density behavior observed as a universal feature of homogeneous holographic configurations \cite{Ecker:2025sjb}. While our approximations are not reliable beyond $3n_S$, this could suggest that the different density dependence reported in Ref.~\cite{Ecker:2025sjb} is primarily an effect of the homogeneous approximation, rather than an intrinsic prediction of the holographic model.

To check the assumption of the lattice calculation, namely a clear separation of core and tail regions, we plot the distance between neighboring instantons in units of twice the radius parameter in the right panel of Fig.~\ref{fig:Lattice}. 
For a typical neutron star, the approximation appears fairly well under control, whereas in the non-BPST case it becomes unjustifiable around $n\gtrsim 5 n_S$, which affects the heaviest neutron stars (their maximal densities are marked by dots in the right panel of Fig.~\ref{fig:Lattice}).

\section{Beta equilibrium, hybrid EOS and NS phenomenology}
\label{sec:betaeq}

To connect the symmetric-matter EOS to realistic dense matter, isospin-breaking corrections need to be accounted for. To this end, we introduce the symmetry energy $S(n_B)$ in the parabolic approximation. The total baryonic energy density is thus given in terms of the proton and neutron densities $n_p$ and $n_n$ as
\begin{align}
\label{eq:energyfull}
    \frac{E(n_B,\delta)}{N_B}&=\frac{E_{\rm S}(n_B)}{N_B}+ S(n_B)\delta^2,\\
      \mathcal{E}(n_B,\delta)&=\mathcal{E}_{\rm S}(n_B)+n_BS(n_B)\delta^2,\qquad \delta\equiv \frac{n_n - n_p}{n_B},
\end{align}
where $\mathcal{E}_{\rm S}$ denotes the isoscalar energy density taken from the lattice of holographic baryons obtained in the previous sections. 
The symmetry-energy sector is parameterized as
\begin{equation}
    S(n_B)=S_0\left(\frac{n_B}{n_S}\right)^{\gamma_1},
\end{equation}
so that the symmetry-energy slope parameter $L$ is given in terms of $\gamma_1$ and $S$ by
\begin{equation}
    L\equiv 3n_S\left.\frac{\partial S}{\partial n_B}\right|_{n_S}=3S_0\gamma_1.
\end{equation}
A determination of the symmetry energy in the WSS model is in principle possible, but also very challenging, especially when considering localized nuclear matter. 
Since the pressure at saturation density vanishes for symmetric nuclear matter, the isospin-breaking contributions to Eq.~\eqref{eq:energyfull} are particularly important in this regime. However, at higher densities the dominant contribution is expected to arise from the isoscalar sector.

Lacking a first-principles computation of the symmetry energy in the WSS model, we use phenomenological input for $S_0$ and the slope parameter $L$ from nuclear EFT as estimated in Ref.~\cite{Hebeler:2013nza}. This methodology represents a controlled extension of symmetric matter to beta-equilibrated matter and, as we show, matches well to the low density EOS  band around $n_B \sim n_S$.

Another common, but less justified, approach is to instead fit the symmetric-matter EOS (obtained from the model) directly to the low-density phenomenological EOS. 
We examine this fitting strategy separately in appendix~\ref{app:equalfooting}, which also allows for a direct comparison with the previous calculations based on the homogeneous ansatz.
This results in a systematically stiffer EOS, as the baryons are forced to provide the pressure normally given by the leptons at saturation, which then propagates to higher densities.

Here we instead keep symmetric matter sector as determined in Sec.~\ref{sec:symmetricfit}, and use it for the construction of the full EOS under the assumption of a realistic isospin-asymmetric sector whose parameters are taken from phenomenology.

This has the advantage of keeping the symmetric matter fit of the WSS model intact and independent of the details of the isovector and lepton sector. We leave holographic determinations of isospin asymmetry for future work.

The fraction of protons to neutrons in realistic neutron star matter is determined by neutron decay and electron capture processes $n\rightarrow p+e^-+\bar \nu_e$ and $p+e^-\rightarrow n+\nu_e$. The neutrinos are assumed to escape the star, thus appearing only in final states.
Beta equilibrium and the parabolic approximation lead to
\begin{equation}
    \mu_n-\mu_p=\mu_e=\mu_\mu\equiv\mu_l, \qquad \mu_l(n_B)=4S(n_B)\,\delta(n_B),
\end{equation}
while charge neutrality requires
\begin{equation}
    n_p=n_e+n_\mu,\qquad  \text{with} \qquad n_p=\frac{1-\delta}{2}n_B.
\end{equation}
The lepton number densities are
\begin{align}
    n_e(\mu_l) &= \frac{\mu_l^3}{3\pi^2(\hbar c)^3},\\
    n_\mu(\mu_l) &= \Theta(\mu_l-m_\mu)\frac{\left(\mu_l^2-m_\mu^2\right)^{3/2}}{3\pi^2(\hbar c)^3}.
\end{align}
At each $n_B$, the asymmetry $\delta(n_B)$ is obtained by solving
\begin{equation}
    \frac{1-\delta(n_B)}{2}n_B-n_e\left(\mu_e(n_B)\right)-n_\mu\left(\mu_\mu(n_B)\right)=0,
\end{equation}
which is then used to compute the lepton number densities.
The leptonic pressure and energy density are treated as  free Fermi gases. For $l=e,\mu$ with mass $m_l$, chemical potential $\mu_l$ and Fermi momentum $p_{{\rm F} l}=\sqrt{\max(\mu_l^2-m_l^2,0)}$, 
\begin{align}
P_l&=\frac{\Theta(\mu_l-m_l)}{24\pi^2(\hbar c)^3}\left[p_{{\rm F} l}\mu_l\left(2p_{{\rm F} l}^2-3m_l^2\right)+3m_l^4\log\frac{p_{{\rm F} l}+\mu_l}{m_l}\right],\\
\mathcal{E}_l&=\frac{\Theta(\mu_l-m_l)}{8\pi^2(\hbar c)^3}\left[p_{{\rm F} l}\mu_l\left(2p_{{\rm F} l}^2+m_l^2\right)-m_l^4\log\frac{p_{{\rm F} l}+\mu_l}{m_l}\right].
\end{align}
Finally, the isovector baryonic contributions used in the EOS are
\begin{equation}
    \mathcal{E}_{\rm isovec}(n_B)=n_BS(n_B)\,\delta(n_B)^2,
    \qquad
    P_{\rm isovec}(n_B)=\gamma_1 n_BS(n_B)\,\delta(n_B)^2.
\end{equation}
The total energy density and pressure are obtained by adding the leptonic and isovector contributions to the symmetric nuclear matter expression \eqref{eq:energydensitysymm}. Parameters for $S_0$ and $L$ are extracted from Ref.~\cite{Hebeler:2013nza}, where a whole band of possible values are given. For both the BPST and non-BPST symmetric matter construction, this results in an ensemble of EOS, where each element corresponds to one of the allowed values of $S_0,L$ given in Ref.~\cite{Hebeler:2013nza}. 

\begin{figure}
    \centering
    \includegraphics[width=0.49\linewidth]{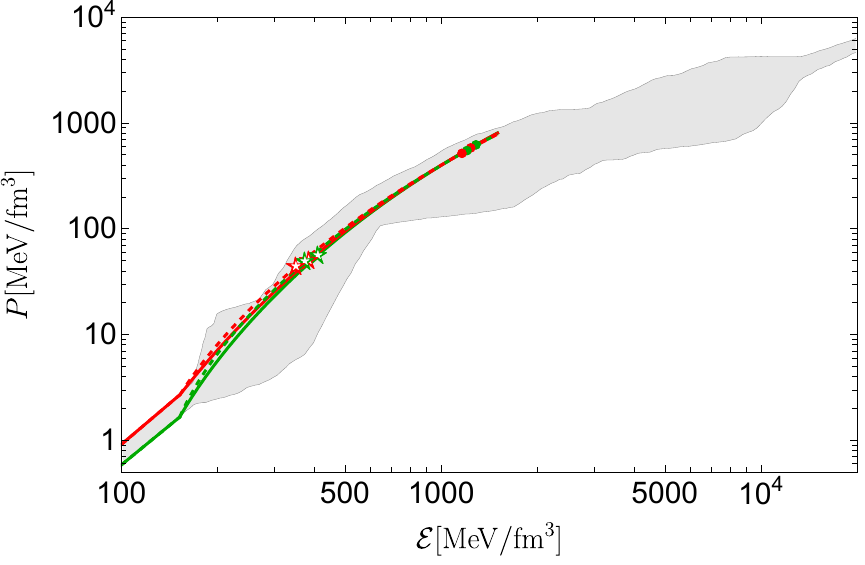}
    \includegraphics[width=0.49\linewidth]{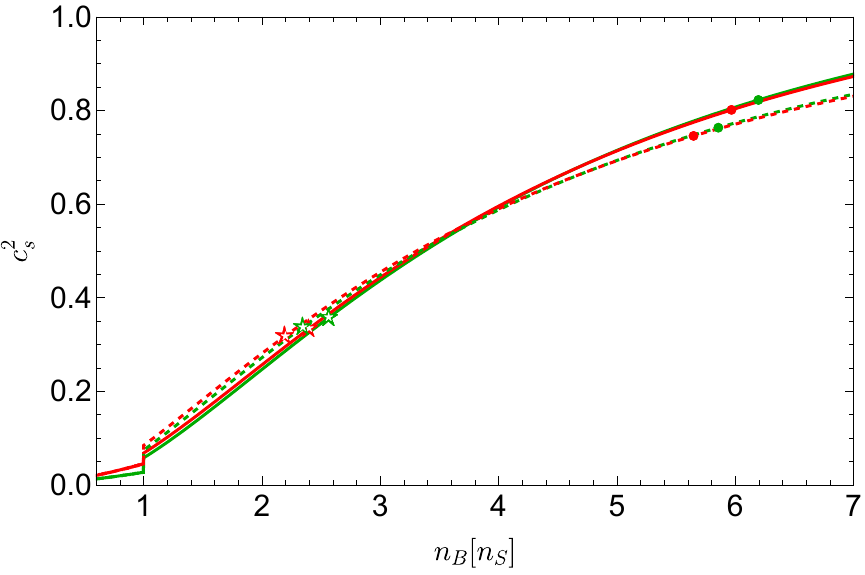}
    \includegraphics[width=0.49\linewidth]{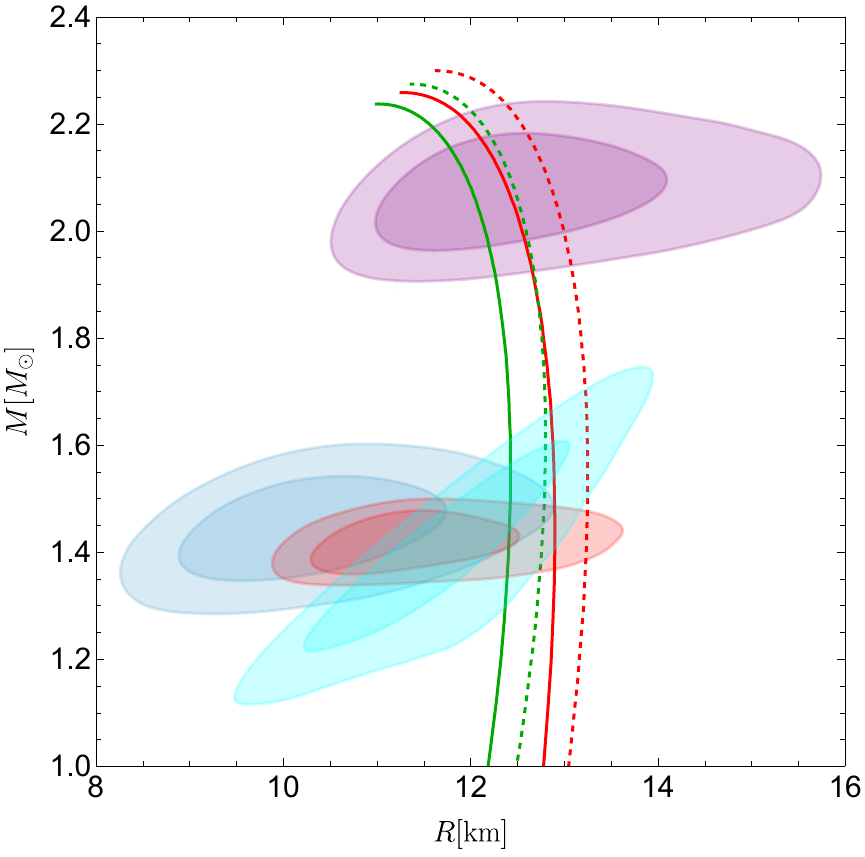}
    \includegraphics[width=0.49\linewidth]{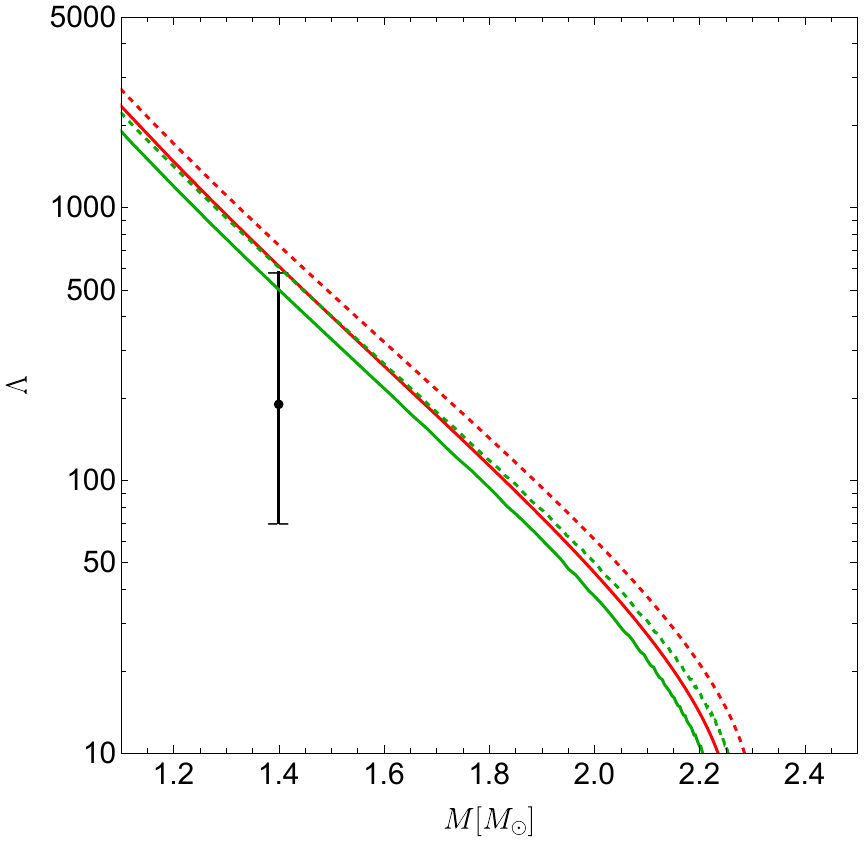}
    \caption{\textbf{Top left:} hybrid equations of state obtained from the calibrated symmetric-matter holographic EOS after introducing the phenomenological symmetry-energy sector. 
     \textbf{Top right:} speed of sound squared as a function of density in units of saturation density for the same hybrid EOS.
    \textbf{Bottom left:} neutron-star mass-radius curves obtained from the hybrid EOS and compared with 68\% and 95\% confidence regions from the latest NICER analyses: PSR J0614-3329 (blue) \cite{Mauviard:2025dmd}, PSR J0740+6620 (purple) \cite{Salmi:2024aum}, PSR J0437-4715 (red) \cite{Choudhury:2024xbk}, PSR J0030+0451 (cyan) \cite{Vinciguerra:2023qxq}.
    \textbf{Bottom right:} tidal deformability as a function of mass for neutron stars built from the hybrid EOS. In every panel, the green and red lines correspond respectively to $(S_0,L)=(29.96,32.62)\,\rm MeV$ and $(S_0,L)=(32.76,57.01)\,\rm MeV$, while the solid (dashed) lines correspond to the BPST (non-BPST) choices for $(\gamma,\eta)$. The filled dots on each curve identify the values at the center of the heaviest stable NS, {the open star symbols to those at the center of 1.4 solar mass neutron star.}
    }
    \label{fig:EOS_cs_RM_Tidal_FCC}
\end{figure}

The stiffest (green) and softest (red) representatives of the ensemble are shown in Fig.~\ref{fig:EOS_cs_RM_Tidal_FCC}.  
The BPST (solid) and non-BPST (dashed) nuclear matter EOS match almost smoothly to EFT calculations at saturation density. Only a small jump in the speed of sound remains, as shown in the right panel of Fig.~\ref{fig:EOS_cs_RM_Tidal_FCC}. The resulting EOS clearly satisfy the constraints indicated by the gray band, which are obtained from a combination of low-energy EOS, pQCD and observational data from neutron star mergers \cite{Annala:2017llu}. This is a noticeable difference to the homogeneous ansatz shown in Fig.~\ref{fig:HAEOS_Polytwopoints}. We compare the EOS obtained from the holographic soliton crystal to the homogeneous ansatz more  thoroughly in appendix \ref{app:equalfooting}.

Below saturation density, the holographic crystal ansatz is unlikely to describe matter appropriately due to the importance of the electromagnetic interactions and the appearance of exotic configurations such as nuclear pasta. In the construction of neutron star matter, we thus simply use nuclear EFT EOS below $n_S$.

Using these ingredients, we construct hybrid EOS $\mathcal{E}(P)$ for neutron-star matter and determine stellar observables by solving the TOV equations for static stars,
\begin{align}\label{eq:TOV1}
  \frac{\d P}{\d r}&= -G(\mathcal{E}+ P)\frac{m+4\pi r^3 P}{r(r-2Gm)},\\
  \frac{\d m}{\d r}&= 4\pi r^2 \mathcal{E},\label{eq:TOV2}
\end{align}
further supplemented by a third equation to calculate the tidal deformability $\Lambda$, defined in terms of the compactness $c=\frac{GM}{R}$ and the tidal Love number $k_2$ as 
\begin{equation}
    \Lambda=\frac{2 k_2}{3 c^5}.
\end{equation}
The tidal Love number is given in terms of a radial function $y(r)$ evaluated at $r=R$, thus denoting $y_R\equiv y(R)$:
\begin{align}
    k_2=&\frac{8c^5}{5}(1-2c)^2[2-y_R +2c(y_R-1)]\times\{2c [ 6-3y_R+3c(5y_R-8)] +\nonumber\\
    &+4c^3[13-11y_R+c(3y_R-2)+2c^2(1+y_R)]+\nonumber\\
    &+3(1-2c)^2[2-y_R+2c(y_R-1)]\ln(1-2c) \}^{-1}.
\end{align}
Finally, the function $y(r)$ can be obtained by solving the differential equation
\begin{align}
    r\frac{dy}{dr}=-y^2-&\frac{4\pi Gr^2\left(5\mathcal{E}+9P+\frac{\mathcal{E}+P}{c_s^2}\right)-6}{1-\frac{2GM}{r}}-\nonumber\\
    &-\frac{y[1-4\pi G r^2\left(\mathcal{E}-P\right)]}{1-\frac{2GM}{r}}+\frac{4G^2\left(M+4\pi P r^3\right)^2}{r^2\left(1-\frac{2GM}{r}\right)^2},
\end{align}
with boundary condition $y(0)=2$.

The resulting mass-radius curves are shown in {the bottom left panel of} Fig.~\ref{fig:EOS_cs_RM_Tidal_FCC} and are consistent with current NICER constraints. For a $1.4 \,M_\odot$ neutron star, central number densities of around $n_B\sim (2.2 -2.6) \, n_S$ are reached. As shown in Fig.~\ref{fig:Lattice} (indicated by small red and green stars), at these densities the cores of the instantons are still reasonably separated and the
soliton core/tail approximation is expected to be trustworthy. 
For heavier neutron stars, and particularly for the most massive ones, the cores of  the instantons start to overlap and the potential \eqref{eq:dinuclpotential} is not appropriate anymore. In this regime, only a full numerical simulation can give insights into the behavior of the model. We also note that a transition to quark matter may occur in the cores of the most massive neutron stars. 
After accounting for backreactions of the flavor-branes on the metric, a quark phase should be present in the (original)\footnote{By working in a version of the WSS model with deconfined geometry and small D8-$\overline{\mathrm{D8}}$ brane separation, transitions between mesonic, baryonic, and quark phases can been arranged without the need of flavor-brane backreactions and have been studied e.g.\ in Ref.~\cite{BitaghsirFadafan:2018uzs}.} WSS model, indications of which have been studied in Ref.~\cite{Bigazzi:2014qsa}. Whether it can accurately describe realistic quark matter is unclear at the moment. (The WSS model is expected to decompactify for energies much larger than $M_{\rm KK}$. Therefore, it necessarily deviates from QCD at high energies, and probably also at large chemical potentials/densities.)

Another interesting observable in the context of neutron stars is the (dimensionless) tidal deformability $\Lambda$. It quantifies the distortion of the shape of a neutron star in the gravitational field of another neutron star during the inspiral and merger periods. We show the results for the different fits in the bottom right panel of Fig.~\ref{fig:EOS_cs_RM_Tidal_FCC}. The predictions of the model sit at the upper part of the experimental estimate $\Lambda(1.4\, M_\odot)=190^{+390}_{-120}$ of Ref.~\cite{LIGOScientific:2018cki}, with the softest constructions still being compatible with it, while the stiffest show some tension.

\section{Conclusions}\label{sec:conclusions}

In this exploratory work we revisited the construction of dense baryonic matter in the Witten-Sakai-Sugimoto model, going beyond the customary homogeneous ansatz by constructing a FCC crystal of baryonic solitons with nearest neighbors placed in the most attractive channel of a two-baryon interaction potential derived along the lines of Ref.~\cite{Baldino:2017mqq, Baldino:2021uie}, which assumes a sufficient separation of core and tail regions of the SS solitons.
Focusing on symmetric nuclear-matter observables, we fixed $(\lambda,M_{\rm KK})$ from saturation-density and onset-chemical-potential properties, and used the quark-mass parameter $m$ to reproduce the physical pion mass.

Curvature corrections to the mass formula of baryons were parametrized by
two sets of numbers $(\gamma,\eta)$.
For the BPST-based choice of $(\gamma,\eta)$, the resulting parameters $\lambda=18.26$ and $M_{\rm KK}=661.5\,\mathrm{MeV}$ are accompanied by $M_B=942.0\,\mathrm{MeV}$ and binding energy per nucleon $-19.3$ $\mathrm{MeV}$ for symmetric matter at saturation, both not too far off from phenomenology while remaining reasonably close to the parameters of the model required for mesonic physics. The joint reductions of $f_\pi$ and $M_{\rho,\omega}$ compared to the standard choice \cite{Sakai:2004cn} actually happen to be roughly in line with expectations from Brown-Rho scaling \cite{Brown:1991kk} in a dense medium, whereas the homogeneous ansatz typically leads to a reduced $f_\pi$ but even somewhat increased vector meson mass.

The incompressibility $K(n_S)=374.9\,\mathrm{MeV}$ 
obtained in our approach with the above parameters remains somewhat high, but represents a clear improvement over the homogeneous ansatz, which overshoots the phenomenological values by about an order of magnitude. The non-BPST choice of $(\gamma,\eta)$ leads to somewhat stronger tension with saturation properties and baryon mass, but still allows for the construction of hybrid EOS compatible with NICER data. This last set of parameters also results in baryons beginning to appreciably overlap at smaller densities compared to the BPST set, and its corresponding value of $\lambda=45.54$ deviates significantly from $\lambda=16.63$ at which the numerical calculation of the baryon mass is computed and around which the rescaling factors $(\gamma,\eta)$ can be considered a good approximation.

To address neutron-star matter, we then supplemented the symmetric holographic EOS with a phenomenological symmetry-energy sector in the parabolic approximation, using representative $(S_0,L)$ values from Ref.~\cite{Hebeler:2013nza}. 
This led to a beta-equilibrated hybrid EOS with neutron-star mass-radius and tidal-deformability predictions that are compatible with NICER constraints.

Several limitations remain. The lattice description relies on a truncated interaction sum and on a potential derived from asymptotic soliton tails, and it assumes a specific crystalline structure and channel assignment. 
Furthermore, curvature corrections are implemented by simple scaling factors that are assumed to be independent of the 't Hooft coupling, $\lambda$.
Including the effects of the finite size baryon core, incorporating the isovectorial sector from top-down holography and exploring the effects of temperature and external magnetic field on the lattice are all natural next steps.
A more challenging but very relevant task would be an assessment of $(N\ge3)$-body forces between the holographic solitons.
While the 3-body force is expected to have large consequences, much of the stiffness that it would induce in the EOS will most likely be absorbed into the calibration of the model, when calibrating to phenomenological values at or near saturation density. 
Finally, there may be phase transitions to half-instantons (or in 3D half-Skyrmions) \cite{Rho:2009ym} or phase transitions to quark matter: while not directly relevant for the purpose of describing realistic holographic nuclear matter around saturation or for the properties of the most common neutron stars (for which we find central densities to be always lower than $3 n_S$),
the inclusion of these phases represent also a most interesting
aim for further studies within holographic QCD, and are crucial for a reliable holographic determination of the heaviest possible neutron star mass.

\acknowledgments

We would like to thank Andreas Schmitt for very useful discussions, and we also thank Albino Perego and Domenico Logoteta for sharing their incompressibility results for Fig.~\ref{fig:Binding_K}.
This work has been supported by the Austrian Science Fund FWF, Grant-DOI
\href{https://doi.org/10.55776/PAT7221623}{10.55776/\break PAT7221623}. The work of L.~B.~is also supported by the National Natural Science Foundation of China Youth Grant (Grant no.~12405084).
S.~B.~G.~thanks the Outstanding Talent Program of Henan University for partial support.

\appendix
\section{Homogeneous-ansatz vs FCC lattice hybrid EOS: a direct comparison}
\label{app:equalfooting}

In this appendix we exhibit a detailed comparison of the EOS obtained from the homogeneous ansatz and from the soliton lattice.  

In the main text, it was pointed out that the FCC holographic soliton lattice matches much better with the low-density chiral EFT band than the homogeneous ansatz. The latter EOS was obtained by matching the isospin-symmetric results, obtained in the smeared approximation at a given $\lambda$, to the low density EFT EOS and requiring continuity of energy density, pressure, and baryon number density. By contrast, the isospin-symmetric FCC soliton lattice EOS was first fitted to properties of symmetric nuclear matter, and subsequently endowed with phenomenologically determined isospin-breaking corrections. The comparison of the homogeneous ansatz to the FCC lattice, as done in the main text,  is thus not completely faithful. The purpose of this appendix is to adopt the same fitting and matching procedure for both approaches, allowing for a more accurate comparison.

\begin{figure}[ht]
    \centering
    \includegraphics[width=0.75\linewidth]{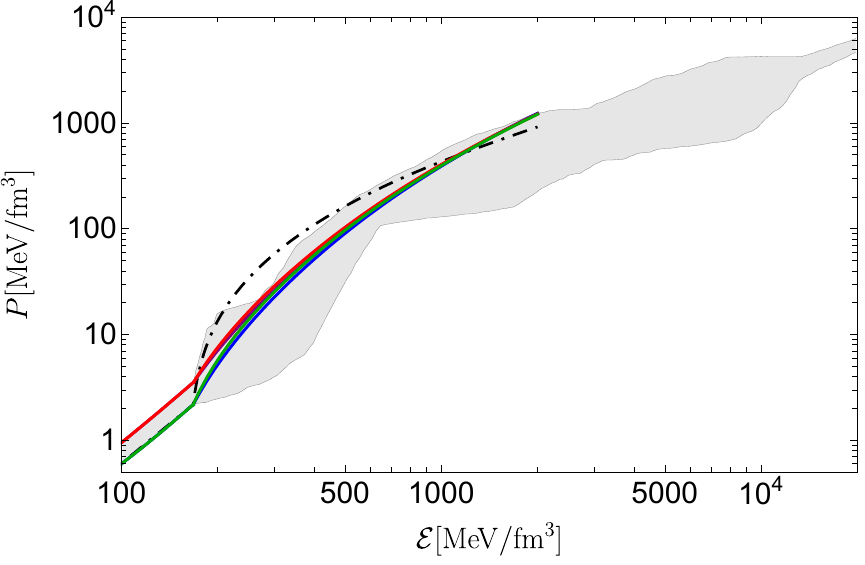}
    \caption{Comparison of matched hybrid EOS built on equal footing with transition at $n_t=1.1\,n_S$ and continuity of $P$, $\mathcal{E}$ and $n_B$. Color coding: the dot-dashed curve corresponds to the homogeneous ansatz; the green (red) curve corresponds to the soft (stiff) non-BPST FCC-lattice matched EOS, the blue (purple) curve corresponds to the soft (stiff) BPST FCC-lattice matched EOS. The homogeneous-ansatz EOS (softest construction with the same matching prescription) becomes too stiff and leaves the allowed region, while FCC-based matched EOS remain compatible with observational constraints.}
    \label{fig:ComparisonHLPSMatch}
\end{figure}

To this end, we match the holographic EOS to the softest nuclear EFT EOS at $n_t=1.1\, n_S$ and require continuous pressure, energy density, and baryon number density.
These conditions fully determine the parameters $\lambda,M_{\rm KK}$ in the EOS of the homogeneous ansatz.
For the FCC lattice, we use both the BPST and non-BPST values for $(\gamma,\eta)$ and find $\lambda, M_{\rm KK}$ with the same procedure. The results are shown in Fig.~\ref{fig:ComparisonHLPSMatch}. Additionally, we also plot EOS of the FCC lattice matched to the stiffest nuclear EFT EOS at $1.1 \, n_S$. 

The parameters obtained in this procedure read:
\begin{itemize}
    \item soft BPST $(\lambda,M_{\rm KK})=(22.6,624.3\,\mathrm{MeV})$,
    \item stiff BPST $(\lambda,M_{\rm KK})=(25.4,592.8\,\mathrm{MeV})$,
    \item soft non-BPST $(\lambda,M_{\rm KK})=(50.1,443.1\,\mathrm{MeV})$,
    \item stiff non-BPST $(\lambda,M_{\rm KK})=(52.7,428.2\,\mathrm{MeV})$.
\end{itemize} For the homogeneous ansatz, matching to the the softest EFT construction gives $(\lambda,M_{\rm KK})=(7.09,1033.3\,\mathrm{MeV})$.\footnote{In contrast to the FCC-lattice cases, the homogeneous ansatz corresponds to a reduction of $f_\pi$ together with an increase of $M_{\rho,\omega}$, contrary to the prediction of an in-medium reduction of both according to Brown-Rho scaling \cite{Brown:1991kk}.}

 Even under these modified matching conditions, the FCC-lattice construction remains compatible with observational EOS constraints. The homogeneous ansatz leads, under the assumed continuity of pressure and number density, to comparatively stiffer EOS incompatible with the observed properties of nuclear matter at saturation.

We interpret this as evidence that the lattice construction with the associated approximate interaction potential \eqref{eq:dinuclpotential} is a better description of nuclear matter around saturation density, while the smeared approximation fails to capture the correct physics in this regime. However, since both approaches involve approximations, one should not  take this as a definitive statement. Ultimately, only a numerical simulation of dense baryonic configurations in the WSS model can provide more accurate information.

\section{FCC shell counting}
\label{app:FCC-shell-counting}

In this appendix we explicitly show the shell-counting procedure used for the FCC crystal.
The nearest-neighbor distance is denoted by $R$. It is convenient to parametrize all FCC lattice vectors relative to a reference baryon as
\begin{equation}
    \vec r =
    \frac{R}{\sqrt2}(u,v,w),
    \qquad
    u,v,w\in\mathbb Z,
    \qquad
    u+v+w\in 2\mathbb Z .
\label{eq:FCC-lattice-vectors}
\end{equation}
The squared distance is then
\begin{equation}
    |\vec r|^2
    =
    \frac{R^2}{2}(u^2+v^2+w^2)
    \equiv nR^2,
    \qquad
    n\in\mathbb N .
\label{eq:FCC-shell-index}
\end{equation}
Thus a spherical shell is labeled by the integer
\begin{equation}
    n=\frac{u^2+v^2+w^2}{2},
    \qquad
    |\vec r|=\sqrt n\,R .
\end{equation}
The shell multiplicity is obtained by counting all integer triples that satisfy
Eq.~\eqref{eq:FCC-lattice-vectors} and Eq.~\eqref{eq:FCC-shell-index}.

The FCC orientation pattern is fixed by requiring all nearest-neighbor interactions to lie in the attractive channel. The twelve nearest-neighbor vectors are
\begin{equation}
    \vec r =
    \frac{R}{\sqrt2}(\pm1,\pm1,0),
    \qquad
    \text{and permutations},
\end{equation}
and we assign the relative orientations
\begin{equation}
\begin{split}
    B^\dagger C &= \pm \i\sigma_3,
    \qquad
    \vec r=\frac{R}{\sqrt2}(\pm1,\pm1,0),\\
    B^\dagger C &= \pm \i\sigma_1,
    \qquad
    \vec r=\frac{R}{\sqrt2}(0,\pm1,\pm1),\\
    B^\dagger C &= \pm \i\sigma_2,
    \qquad
    \vec r=\frac{R}{\sqrt2}(\pm1,0,\pm1).
\end{split}
\label{eq:FCC-nearest-orientations}
\end{equation}
Equivalently, the relative orientation is a rotation by $\alpha=\pi$ around the coordinate axis corresponding to the vanishing entry of the nearest-neighbor displacement.
This prescription extends consistently to the full FCC lattice. If $n$ is even, all three integers $(u,v,w)$ are even, and the relative orientation belongs to the trivial class,
\begin{equation}
    n\ \text{even}:
    \qquad
    B^\dagger C = \mathds 1 ,
    \qquad
    \alpha=0 .
\end{equation}
If $n$ is odd, exactly one of $(u,v,w)$ is even and the other two are odd. The relative orientation is then a rotation by \(\alpha=\pi\) around the axis selected by the unique even
entry:
\begin{equation}
    n\ \text{odd}:
    \qquad
    B^\dagger C = \pm \i\sigma_j ,
    \qquad
    j=
    \begin{cases}
        1, & u\ \text{even},\\
        2, & v\ \text{even},\\
        3, & w\ \text{even}.
    \end{cases}
\label{eq:FCC-odd-shell-axis}
\end{equation}
The sign is irrelevant for the two-body potential, since the orientation enters through
\begin{equation}
    M_{ij}(G)
    =
    \frac12{\rm tr}\left(\sigma_iG\sigma_jG^\dagger\right),
    \qquad
    G=B^\dagger C .
\end{equation}

For the $\alpha=0$ class, all sites in a fixed shell contribute equally. For the $\alpha=\pi$ class, the contraction $M_{ij}P_{ij}$ depends on the {inner product $\hat{u} \cdot r$ of the separation vector $r$ and the rotation axis $\hat{u}$ in flavor space. Denoting $\hat{u} \cdot r=r_e R/\sqrt{2}$, the variable $r_e$ is always an integer and one obtains}
\begin{equation}
    \frac{(\hat u\cdot \vec r)^2}{r^2}
    =
    \frac{r_e^2}{2n}.
\label{eq:FCC-projection-factor}
\end{equation}
Therefore odd shells must be further split into orientation-projection subclasses labeled
by
\begin{equation}
    s=r_e^2.
\end{equation}
{For nearest neighbors, $s=0$, but starting from the shell $n=3$, nonzero values of $r_e^2$ are possible. The first shell, where two different values of $s$ appear, is $n=9$.}
We denote by $N_n^{(0)}$ the number of sites in shell $n$ with $\alpha=0$, and by $N_{n,s}^{(\pi)}$ the number of sites in shell $n$ with $\alpha=\pi$ and $r_e^2=s$.
With this notation, the one-baryon energy entering the FCC equation of state can be
written as
\begin{equation}
    E_p(R)
    =
    M_B
    +
    \frac12
    \sum_{n=1}^{n_{\rm max}}
    \left[
        N_n^{(0)} V_0(\sqrt n R)
        +
        \sum_s
        N_{n,s}^{(\pi)}
        V_\pi\!\left(\sqrt n R;\frac{s}{2n}\right)
    \right],
\label{eq:FCC-shell-sum}
\end{equation}
where $V_0$ is the two-body potential evaluated with $B^\dagger C=\mathds 1$, while
$V_\pi(r;\chi)$ denotes the potential evaluated with $\alpha=\pi$ and
$(\hat u\cdot\vec r)^2/r^2= \frac{r_e^2}{2n}$. The factor $1/2$ compensates for double counting of pairs.
The cutoff $n_{\rm max}$ is selected to be high enough for the sum to converge: the shell multiplicities up to $n=30$ are listed in Table~\ref{tab:FCC-shells}. Empty entries correspond to absent shells. 

\begin{table}[t]
\centering
\begingroup
\footnotesize
\setlength{\tabcolsep}{3pt}
\resizebox{0.85\textwidth}{!}{
\begin{tabular}{cc@{\hspace{1.6em}}cc}
\hline
\(n=(r/R)^2\) &  \multicolumn{1}{c}{multiplicities} & \(n=(r/R)^2\) &  \multicolumn{1}{c}{multiplicities} \\
\hline
1             & \(N_{1,0}^{(\pi)}=12\) &    16             & \(N_{16}^{(0)}=12\) \\
2         & \(N_{2}^{(0)}=6\) &    17      & \(N_{17,0}^{(\pi)}=24,\quad N_{17,16}^{(\pi)}=24\) \\
3          & \(N_{3,4}^{(\pi)}=24\) & 18        & \(N_{18}^{(0)}=30\) \\
4             & \(N_{4}^{(0)}=12\) & 19      & \(N_{19,4}^{(\pi)}=48,\quad N_{19,36}^{(\pi)}=24\) \\
5           & \(N_{5,0}^{(\pi)}=24\) & 20       & \(N_{20}^{(0)}=24\) \\
6          & \(N_{6}^{(0)}=8\) & 21      & \(N_{21,16}^{(\pi)}=48\) \\
7          & \(N_{7,4}^{(\pi)}=48\) & 22       & \(N_{22}^{(0)}=24\) \\
8        & \(N_{8}^{(0)}=6\) & 23       & \(N_{23,36}^{(\pi)}=48\) \\
9              & \(N_{9,0}^{(\pi)}=12,\quad N_{9,16}^{(\pi)}=24\) & 24      & \(N_{24}^{(0)}=8\) \\
10     & \(N_{10}^{(0)}=24\) & 25         & \(N_{25,0}^{(\pi)}=36,\quad N_{25,16}^{(\pi)}=48\) \\
11      & \(N_{11,4}^{(\pi)}=24\) & 26    & \(N_{26}^{(0)}=24\) \\
12       & \(N_{12}^{(0)}=24\) & 27       & \(N_{27,4}^{(\pi)}=72,\quad N_{27,36}^{(\pi)}=24\) \\
13       & \(N_{13,0}^{(\pi)}=24,\quad N_{13,16}^{(\pi)}=48\) & 28       & \(N_{28}^{(0)}=48\) \\
14      & absent & 29      & \(N_{29,0}^{(\pi)}=24\) \\
15       & \(N_{15,4}^{(\pi)}=48\) & 30      & absent \\
\hline
\end{tabular}
}
\endgroup
\caption{
FCC shell multiplicities up to $n=30$. Even shells have trivial relative orientation
and are counted by $N_n^{(0)}$. Odd shells have $\alpha=\pi$ and are decomposed into
projection subclasses $N_{n,s}^{(\pi)}$, where $s=r_e^2$ is the square of the projected
integer component selected by the unique even entry in $(u,v,w)$. The corresponding projection factor entering
$M_{ij}P_{ij}$ is $s/(2n)$.
}
\label{tab:FCC-shells}
\end{table}

\section{Impact of lattice geometry: results for the simple cubic lattice}
\label{app:latticegeometry}
\begin{figure}
    \centering
    \includegraphics[width=0.40\linewidth]{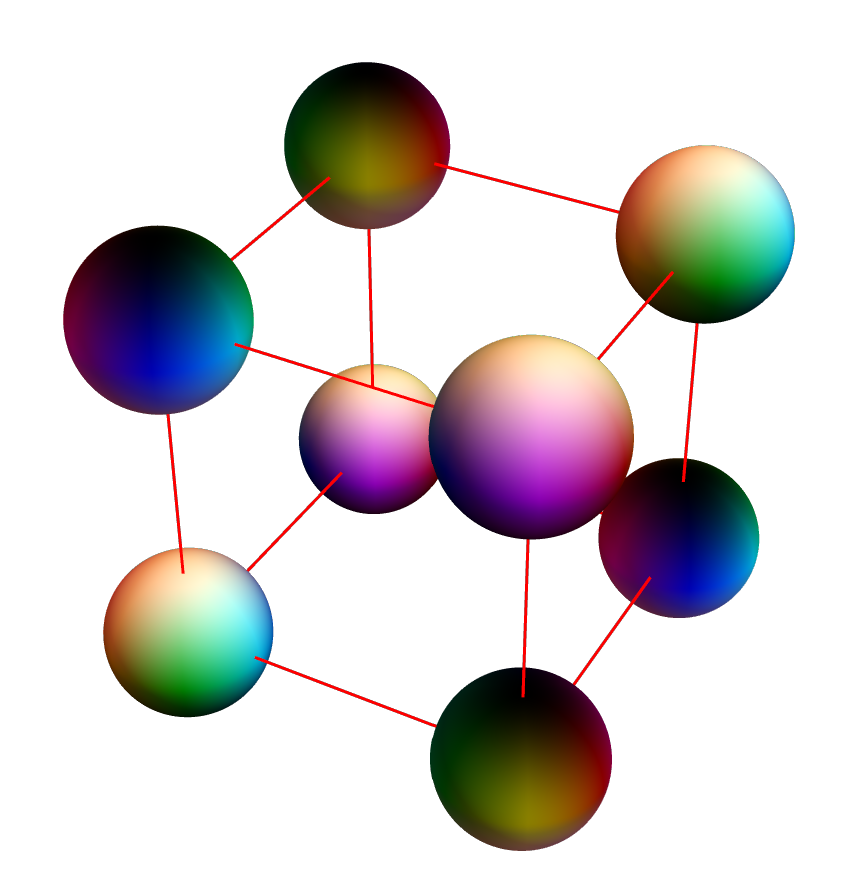}
        \includegraphics[width=0.58\linewidth]{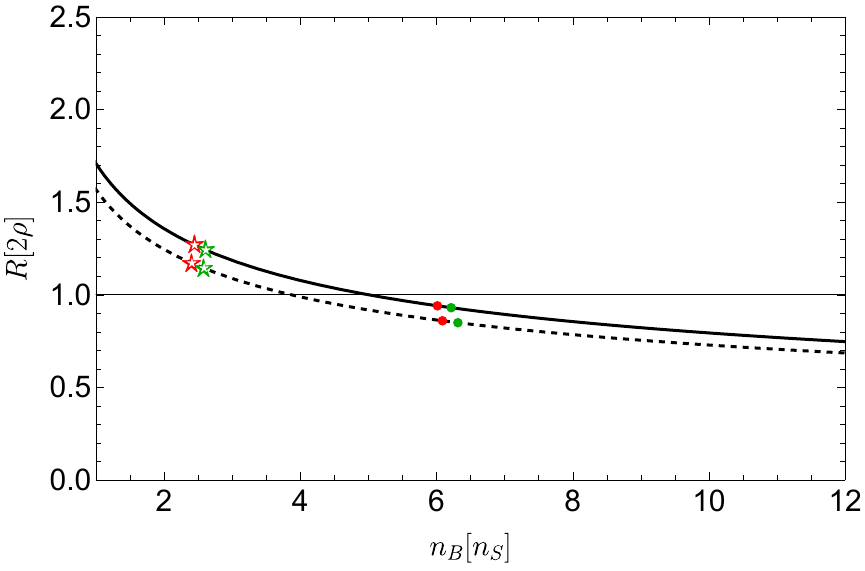}
    \caption{\textbf{Left:} The unit cell of the Klebanov lattice (simple cubic (SC)). The unit cell contains $B=1$ baryons, with each corner contributing $1/8$ of the total baryon number of the cell. \textbf{Right:} Separation between the solitons in units of their diameter. Compared to the FCC geometry (Fig.~\ref{fig:Lattice}), the approximation of well separated instanton cores breaks down at lower density as a consequence of the less efficient packing of the SC lattice.}
    \label{fig:SkyrmionPlotSCUnit}
\end{figure}
In this appendix we study the effect of the lattice configuration on the equation of state, focusing specifically on the simple cubic (SC) lattice
considered originally by Klebanov \cite{Klebanov:1985qi}. The FCC lattice, corresponding to densest packing in three dimensions, is generally energetically preferred over the SC lattice in Skyrme models. In the present holographic model, the conclusions are similar. 

At higher densities, transitions to more exotic structures such as the half-instanton lattice may also occur in the WSS model. 
The comparison to the SC lattice therefore also serves the purpose of estimating the expected changes in the EOS and the corresponding neutrons stars, that occur in such transitions. We find that the impact of the lattice geometry is generally rather minimal after including the effects of isospin breaking phenomenologically.
This suggests that the FCC lattice provides robust predictions for the EOS and neutron star observables in the regime where the approximations, discussed in the main text, are appropriate.

The SC lattice, represented graphically in the left panel of Fig.~\ref{fig:SkyrmionPlotSCUnit}, contains one baryon per unit cell, and its sites are parametrized as
\begin{equation}
    \vec r=R\vec m,\qquad 
    \vec m=(m_1,m_2,m_3)\in\mathbb Z^3 ,
\end{equation}
where $R$ is the nearest-neighbor distance. The baryon number density is therefore
\begin{equation}
    n_B^{\rm SC}=\frac{1}{R^3}.
\end{equation}

\begin{figure}
    \centering
    \includegraphics[width=0.49\linewidth]{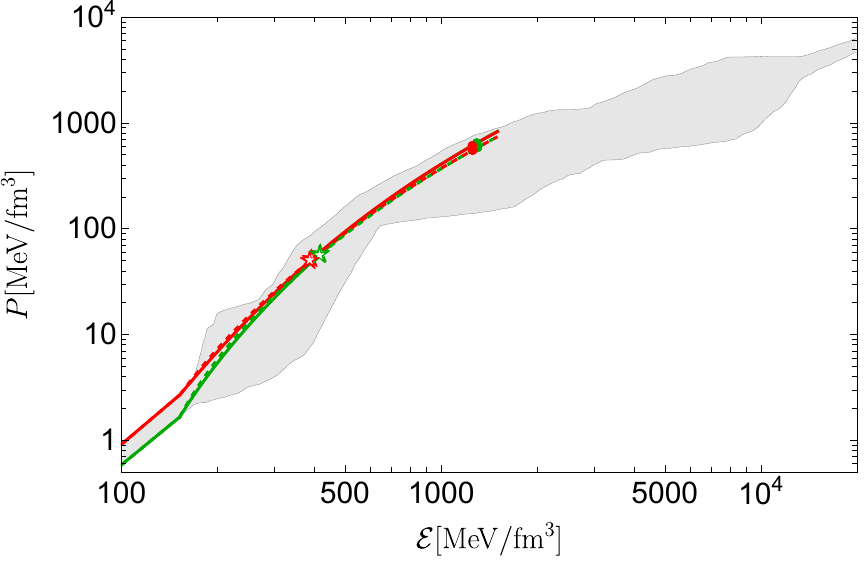}
    \includegraphics[width=0.49\linewidth]{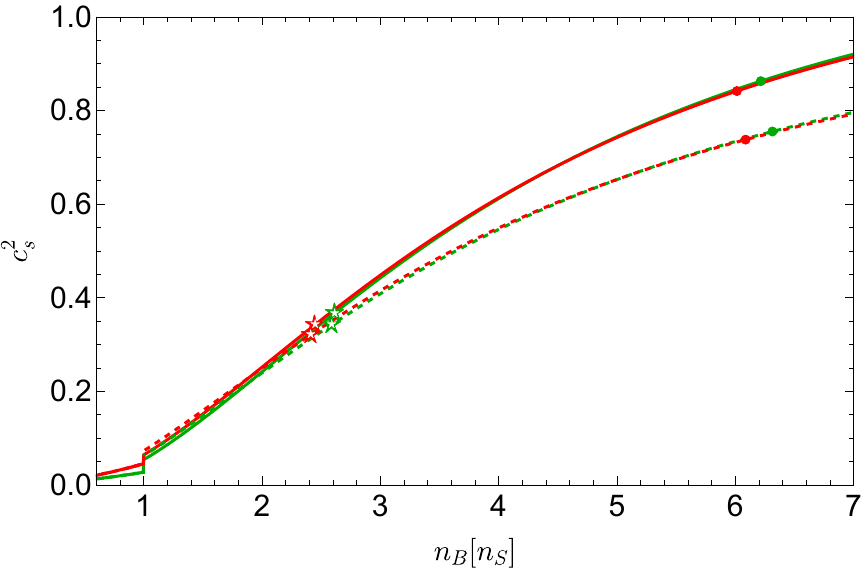}
    \includegraphics[width=0.49\linewidth]{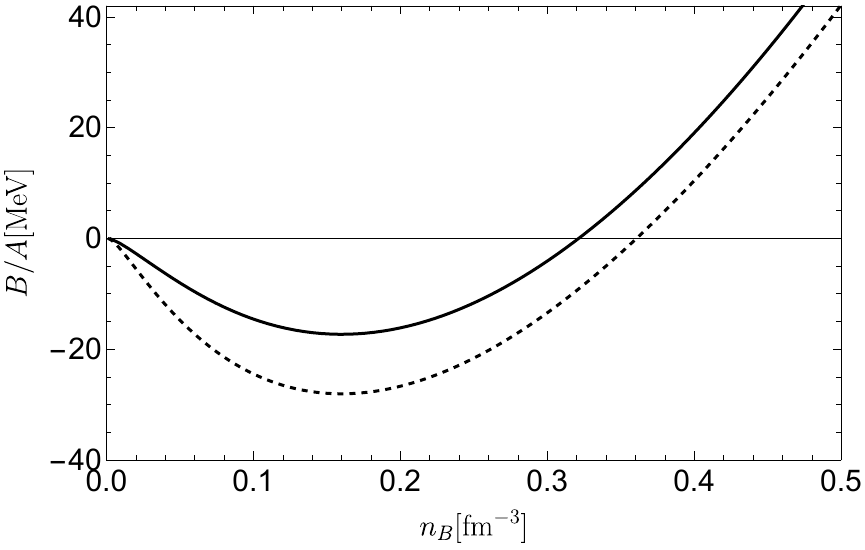}
    \includegraphics[width=0.49\linewidth]{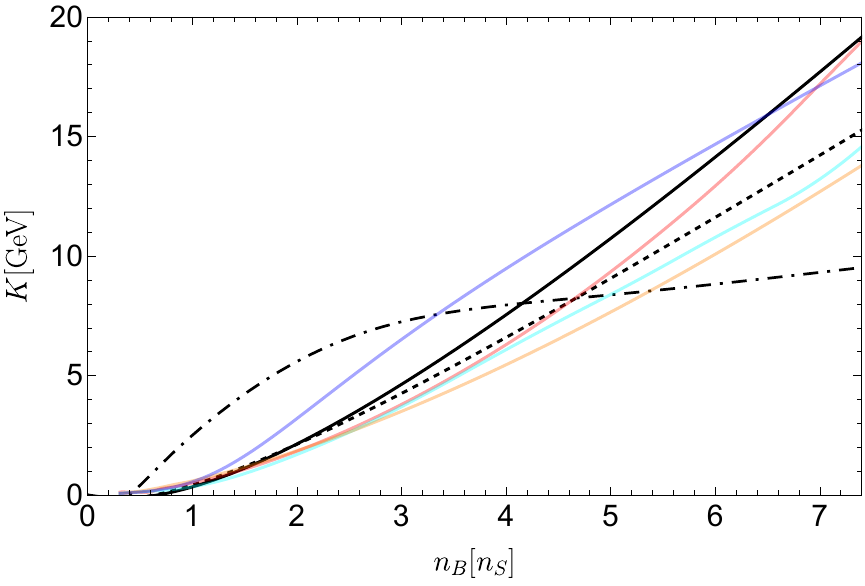}
    \caption{\textbf{Top left:} hybrid equations of state obtained for the SC-lattice-derived symmetric matter and then extended with the phenomenological isovectorial sector as done for the FCC lattice. 
     \textbf{Top right:} Speed of sound squared as a function of density for the hybrid EOS plotted in the left panel.
    \textbf{Bottom left:} Binding energy per nucleon as a function of baryon number density for the two fits resulting from the BPST (solid) and non-BPST (dashed) choices for parameters $(\gamma,\eta)$.
    \textbf{Bottom right:} Incompressibility as a function of baryon number density for the two fits resulting from the BPST and non-BPST choices for parameters $(\gamma,\eta)$. The dot-dashed black line corresponds to the EOS of the homogeneous ansatz with the same parameters as specified in the caption of Fig.~\ref{fig:Binding_K}. The colored lines show the incompressibility computed in Ref.~\cite{Perego:2021mkd} for a selection of non-holographic models of nuclear matter as listed in the caption of Fig.~\ref{fig:Binding_K}.}
     \label{fig:SC_properties}
\end{figure}
We choose  the baryon at the origin of the coordinate system to have orientation \(B_{\vec 0}=\mathds 1\), and impose that all nearest-neighbor interactions are in the attractive channel. This is achieved by assigning the relative orientations
\begin{equation}
\begin{split}
    B_{\vec 0}^{\dagger}C_{\pm \hat x} &= \pm \i\sigma_2,\\
    B_{\vec 0}^{\dagger}C_{\pm \hat y} &= \pm \i\sigma_3,\\
    B_{\vec 0}^{\dagger}C_{\pm \hat z} &= \pm \i\sigma_1.
\end{split}
\label{eq:SC-nearest-neighbor-orientations}
\end{equation}
We subsequently repeat the procedure after equation \eqref{eq:baryonenergy} of section \ref{sec: nuclear matter} to compute the EOS. Similarly to the FCC crystal, we obtain two different equations of state corresponding to the BPST and non-BPST parameter sets \eqref{eq:ch1} and \eqref{eq:ch2} respectively. As before, we fit $(\lambda,M_{\rm KK})$ to  the onset baryon chemical potential and to  the onset number density $n_S$.
For the BPST fit, we find
\begin{equation}
    \lambda =11.09,\qquad M_{\rm KK}=770.98\rm \,MeV,
\end{equation}
leading to the predictions 
\begin{equation}
    M_B=940.1\, \rm MeV,
\end{equation}
\begin{equation}
    \frac{E}{A} -M_B=-17.4\, \rm MeV,
\end{equation}
\begin{equation}
    K(n_S)=341.3\, \rm MeV.
\end{equation}
 For the non-BPST fit we find instead:
\begin{equation}
    \lambda=34.3,\qquad M_{\rm KK}=531.7\, \rm MeV,
\end{equation}
with the following observables at saturation:
\begin{equation}
    M_B=950.8\, \rm MeV,
\end{equation}
\begin{equation}
    \frac{E}{A} -M_B=-28.1\, \rm MeV,
\end{equation}
\begin{equation}
    K(n_S)=416.2\, \rm MeV.
\end{equation}
The predicted (isospin symmetric) nuclear observables do not differ significantly from the related FCC results of Sec.~\ref{sec:symmetricfit}.
We observe that the symmetric-matter EOS of the SC crystal is generally softer at saturation corresponding to smaller 't Hooft coupling  and larger scale $M_{\rm KK}$. Extrapolation to higher densities shows that the EOS for the BPST parameters then becomes stiffer than the associated FCC derived EOS. However, in the regime where this happens, the approximations implicit in the calculations break down. For the SC lattice, this happens earlier than for the FCC lattice because of the less efficient packing of the baryons. This is indicated in the right panel of Fig.~\ref{fig:SkyrmionPlotSCUnit}.
Interestingly, the phenomenological viability of the lattice-derived EOS at saturation does not depend strongly on the specific lattice geometry. The equation of state, after including isospin-breaking effects as in Sec.~\ref{sec:betaeq} and the speed of sound for the SC crystal can be found in the top left and top right panels of Fig.~\ref{fig:SC_properties}. The binding energy and incompressibility are shown in the bottom left and bottom right panels.

Compared to the results from the homogeneous approximation, even the wrong lattice configuration represents a dramatic improvement, suggesting that the key element is the resolution of the individual baryons.

\bibliographystyle{JHEP}
\bibliography{refs2}

\end{document}